\documentclass[12pt,preprint]{aastex62}

\newcommand\aastex{AAS\TeX}

\received{}
\revised{}
\accepted{}
\submitjournal{ApJ}

\shorttitle{\aastex\ Multi-wavelength Timing Observations of 3C~120}
\shortauthors{Marscher et al.}

\begin{document}

\title{X-ray, UV, and Radio Timing Observations of the Radio Galaxy 3C~120}

\correspondingauthor{Alan P.\ Marscher}
\email{marscher@bu.edu}

\author[0000-0001-7396-3332]{Alan P. Marscher}
\affil{Institute for Astrophysical Research \\
Boston University \\
725 Commonwealth Ave. \\
Boston, MA 02215, USA}

\author{Svetlana G.\ Jorstad}
\affil{Institute for Astrophysical Research \\
Boston University \\
725 Commonwealth Ave. \\
Boston, MA 02215, USA}
\affil{Astronomical Institute \\
St. Petersburg State University \\
Universitetskij Pr.\ 28 \\
Petrodvorets \\
198504 St. Petersburg, Russia}

\author{Karen E.\ Williamson}
\affil{Institute for Astrophysical Research \\
Boston University \\
725 Commonwealth Ave. \\
Boston, MA 02215, USA}

\author{Anne L\"ahteenm\"aki}
\affil{Aalto University Mets\"ahovi Radio Observatory \\
Mets\"ahovintie 114\\
FIN-02540 Kylm\"al\"a \\
Finland}
\affiliation{Aalto University Department of Electronics and Nanoengineering \\
P.O. Box 15500\\
FI-00076 Aalto \\
Finland}

\author{Merja Tornikoski}
\affil{Aalto University Mets\"ahovi Radio Observatory \\
Mets\"ahovintie 114\\
FIN-02540 Kylm\"al\"a \\
Finland}

\author{John M.\ Hunter}
\affil{Institute for Astrophysical Research \\
Boston University \\
725 Commonwealth Ave. \\
Boston, MA 02215, USA}

\author{Katya A.\ Leidig}
\affil{Institute for Astrophysical Research \\
Boston University \\
725 Commonwealth Ave. \\
Boston, MA 02215, USA}

\author{Muhammad Zain Mobeen}
\affil{Institute for Astrophysical Research \\
Boston University \\
725 Commonwealth Ave. \\
Boston, MA 02215, USA}

\author{Rafael J.\ C.\ Vera}
\affil{Aalto University Mets\"ahovi Radio Observatory \\
Mets\"ahovintie 114\\
FIN-02540 Kylm\"al\"a \\
Finland}
\affiliation{Aalto University Department of Electronics and Nanoengineering \\
P.O. Box 15500\\
FI-00076 Aalto \\
Finland}

\author{Wara Chamani}
\affil{Aalto University Mets\"ahovi Radio Observatory \\
Mets\"ahovintie 114\\
FIN-02540 Kylm\"al\"a \\
Finland}
\affiliation{Aalto University Department of Electronics and Nanoengineering \\
P.O. Box 15500\\
FI-00076 Aalto \\
Finland}

\begin{abstract}

We report the results of monitoring of the radio galaxy 3C~120 with the {\it Neil Gehrels Swift Observatory}, Very Long Baseline Array, and Mets\"ahovi Radio  Observatory. The UV-optical continuum spectrum and R-band polarization can be explained by a superposition of an inverted-spectrum source with a synchrotron component containing a disordered magnetic field. The UV-optical and X-ray light curves include dips and flares, while several superluminal knots appear in the parsec-scale jet. The recovery time of the second dip was longer at UV-optical wavelengths, in conflict with a model in which the inner accretion disk (AD) is disrupted during a dip and then refilled from outer to inner radii. We favor an alternative scenario in which occasional polar alignments of the magnetic field in the disk and corona cause the flux dips and formation of shocks in the jet. Similar to observations of Seyfert galaxies, intra-band time lags of flux variations are longer than predicted by the standard AD model. This suggests that scattering or some other reprocessing occurs. The 37~GHz light curve is well correlated with the  optical-UV variations, with a $\sim 20$-day delay. A radio flare in the jet occurred in a superluminal knot 0.14 milliarcseconds downstream of the 43~GHz ``core,' which  places the site of the preceding X-ray/UV/optical flare within the core 0.5-1.3~pc from the black hole. The inverted UV-optical flare spectrum can be explained by a nearly mono-energetic electron distribution with energy similar to the minimum energy inferred in the TeV $\gamma$-ray emitting regions of some BL~Lacertae objects.

\end{abstract}

\keywords{accretion: accretion disks --- galaxies: active --- 
galaxies: individual (3C~120) --- radio continuum: galaxies ---
ultraviolet: galaxies  --- X-rays: galaxies }

\section{Introduction} \label{sec:intro}

The connection between the accretion disk (AD) and the jets of active galactic nuclei 
(AGN) is a key aspect of our physical picture of how these extremely energetic objects 
operate. There are, however, only a small number of AGN with luminous, relativistic jets
where emission from the inner AD plus corona of hot electrons is not overwhelmed by
nonthermal emission from one of the two jets. One of these is the Fanaroff-Riley 1
\citep[FR 1][]{FR1974} radio galaxy
3C~120 ($z=0.033$). The mass of the central black hole in 3C~120 as determined from 
time delays between variations in the optical continuum and 
emission-line fluxes (``reverberation mapping''), combined with the virial theorem, is
$5.6{\pm0.5}\times 10^7~M_\odot$ \citep{Bentz2015}, This mass is similar to that 
obtained from comparing the break in the X-ray power spectral density to that of Cygnus 
X-1 \citep{Marshall2009,Chatterjee2009}. Previous {\it Rossi X-ray Timing Explorer} and 
Very Long Baseline 
Array (VLBA) observations of 3C~120 have revealed that the emergence of an apparently 
superluminal radio knot in the jet is preceded by a dip in the X-ray flux \citep{Marscher2002,Chatterjee2009}. Persistent Fe K$\alpha$ emission lines
in the X-ray spectrum of 3C~120 \citep[e.g.][]{Eracleous2000,Ogle2005}
imply that the bulk of its X-ray emission originates in the AD-corona system.
Observations of nine such dip-ejection events and a similar X-ray spectrum in the FR~2 
radio galaxy 3C~111 \citep{Chatterjee2011} suggest that the behavior is a general 
characteristic of radio galaxies.

At radio frequencies, 3C~120 possesses blazar-like characteristics: variability on 
timescales of weeks to years and a one-sided parsec-scale jet with bright superluminal 
knots ejected 2-6 times per year at apparent speeds of 3-9$c$ \citep{Gomez2001,Jorstad2005,Chatterjee2009,Casadio2015,Jorstad2017}. The mean time delay 
of $68\pm14$ days between X-ray dips and the passage of superluminal knots 
through the 43 GHz ``core'' (nearly unresolved, bright emission feature at the upstream
end of the jet in 43 GHz VLBA images) in 3C~120 indicates that the core lies $\sim 0.5$-1 
pc from the corona \citep{Marscher2002,Chatterjee2009} and is probably a standing 
shock in the jet.

While the temporary drop in X-ray flux followed by the appearance of a superluminal radio 
knot in the jet suggests that radio galaxies and black-hole X-ray binary systems
\citep{Fender2004} behave in a similar fashion,
there are important differences, as noted by \citet{Chatterjee2009} and
\citet{Punsly2015}. The inner AD surrounding a supermassive black hole, as in 3C~120, 
is not hot enough to emit X-rays. Rather, the emission from the AD is at UV and
longer wavelengths. The primary X-rays therefore originate in the corona of hot electrons,
which scatters optical-UV photons emitted by the AD up to X-ray energies
\citep[see, e.g.,][]{Fabian2015,Fabian2017}. In the standard model,
some of these X-rays penetrate into the AD, where they are re-processed to
produce the ``reflection'' spectrum that includes the Fe K$\alpha$ line
\citep[e.g.,][]{Ogle2005,Lohfink2013} and UV-optical continuum 
\citep[e.g.,][]{McHardy2014,Gardner2017}.

Two distinct models have emerged to explain the observed X-ray/radio link in 3C~120. 
\citet{Lohfink2013} have suggested that the jet-disk connection occurs via a “jet cycle”
similar to that of X-ray binary systems. The cycle consists of four stages: (1) full AD; 
(2) disappearance of the innermost regions of the AD and ejection of matter out of the 
disk plane, causing an X-ray and UV-optical dip; (3) mass-energy ejection along the jet, 
resulting in a radio flare and the formation of a new superluminal knot; and (4) refilling 
of the AD from outer to inner radii. Under a model by \citet{Chatterjee2009}, the bulk of 
the optical emission in 3C 120 is generated in the AD, while most of the X-ray continuum 
is produced by scattering in the corona of mainly UV radiation emitted by the AD. In this 
model, the corona is actually the base of the jet, very close to the black 
hole, as proposed for X-ray binaries \citep{Markoff2005} and AGN \citep{King2017}.
An increase in the speed of 
flow through the base lowers the density of Compton-scattering electrons, causing a dip in
the X-ray flux while sending a shock wave down the jet. The shock becomes visible in VLBA 
images $~\sim2$ months later as a superluminal knot emerging from the core.

The \citet{Lohfink2013} model predicts that the UV continuum should drop along with the 
X-rays during a flux dip, and then recover first at longer and then shorter wavelengths.
The decrease in UV flux should occur slightly ahead of the start of the X-ray decline. 
Under the \citet{Chatterjee2009} model, the UV-optical flux should decline after the 
X-ray dip starts, if the temperature of the UV-optical emitting region changes in 
response to the lower X-ray flux. In order to test these predictions,
we carried out a nine-month campaign of monitoring observations of 3C 120 with the
X-Ray Telescope (XRT) and UV-Optical Telescope (UVOT) of the Earth-orbiting {\it Neil 
Gehrels Swift Observatory} (hereafter, {\it Swift}). In order to monitor
changes in the relativistic jet, we imaged 3C~120 with the Very Long
Baseline Array (VLBA) at 43 GHz as part of the VLBA-BU-BLAZAR program \citep{Jorstad2016}.
Here we present and interpret the resulting multi-band X-ray, UV, and optical light 
curves, at a cadence of 2-3 observations per week, as well as roughly monthly VLBA images, 
of 3C~120. Section \ref{sec:observations} describes the observations and data reduction
procedures, \S\ref{sec:results} presents the results of the observations,
and \S\ref{sec:discussion} discusses and interprets the results. We summarize our 
conclusions in \S\ref{sec:summary}.

\section{Observations and Data Analysis} \label{sec:observations}

\subsection{X-ray} \label{subsec:xray}

{\it Swift} observed 3C~120 approximately three times per week between 2016 July 14 and 
2017 March 27 with the X-Ray Telescope (XRT) over a photon energy range of 0.3$-$10 keV. 
Most of the observations were in Window Timing mode in 
order to avoid pile-up of photons in individual pixels in between data read-outs.
We processed the data with the standard \texttt{HEAsoft} \citep{Arnaud1996} package
[versions 6.19 (2016 data) and
6.21 (2017 data)], using the task \texttt{xrtpipeline} to identify times of acceptable 
data quality and to calibrate the data. This included examination of the path of 3C~120 
across the detector, which led to 
the exclusion of a small number of observations during which 3C~120 was outside the field 
over $\geq50\%$ of the exposure time. We created an ancillary response file with 
correction for the point-spread function via task \texttt{xrtmkarf}, and then rebinned the 
data with task \texttt{grppha} in order to include at least 20 photons in each energy 
channel.  We used \texttt{XSPEC} \citep{Arnaud1996}
to fit the spectra with a 
single power-law model, setting the neutral hydrogen column density to a fixed 
Galactic value of $1.11\times10^{21}$ cm$^{-2}$ \citep{Dickey1990}. We employed the 
Monte-Carlo method in \texttt{XSPEC} to determine the goodness of the fit at each epoch.
A $\chi^2$ per
degree of freedom less than 2 was required to consider the fit to be acceptable.
Uncertainties in the spectral index, normalization, and integrated flux are given at 
the 90\% confidence level.

\citet{Lohfink2013} fit the X-ray spectrum of 3C~120 with a model that includes
a power-law soft-excess component. We can place a limit on the level of such a
soft excess in our {\it Swift} spectra by adding a second power-law component and
determining how high its flux could be without degrading the $\chi^2$ of the fit to the data to a statistically unacceptable value. \citet{Lampton1976} determined that, for three
adjustable parameters, an increase in the $\chi^2$ by 11.3 from the best-fit value allows
one to reject a spectral model at the 99\% confidence level. We fit the X-ray spectra 
from four epochs during relatively quiescent periods of the X-ray light curve with the
double power-law model. We set the power-law photon index of the tentative soft-excess 
spectrum to 2.5 --- the lowest value derived for the \citet{Lohfink2013} spectral fits
(which minimizes the disagreement with our data) --- and allowed the spectral index of the 
main power-law component, as well as the flux normalizations of both components, to vary. 
Adding the soft excess component did not improve the fit at 
any of the epochs. We found that the maximum allowed normalization of the soft excess is
$3\times10^{-3}$ photons keV$^{-1}$ cm$^{-2}$ s$^{-1}$, about half of the lowest value in 
the \citet{Lohfink2013} model.

\subsection{Ultraviolet and Optical} \label{subsec:ouv}

Simultaneously with the XRT observations, fluxes were measured in the UVW2 (central
wavelength of 192.8 nm), UVM2 (224.6 nm), UVW1 (260.0 nm), U (346.5 nm), and V (546.8 nm) 
bands of the UV-Optical Telescope ({\it UVOT}) on board {\it Swift}. We incorporated
the current (at the time of data processing) version of the calibration files with
the current version of the HEAsoft package 
to reduce the data. Fluxes were extracted over an aperture of radius $5''$ after 
subtraction of the background flux measured within a source-free region of radius $20''$. 
We summed all exposures composing each image with \texttt{uvotimsum}, and processed 
the data with \texttt{uvotsource} with parameter $\sigma = 5$. Five of the UVOT fluxes
thus derived were isolated outliers, with the UVW2 flux lower than contiguous points by 
30-50\% and fluxes at the other bands lower by smaller fractions. Similar
outliers have been investigated by \citet{Gelbord2015}, who determined that they are
caused by bad pixels on the CCD. We therefore have excluded the outliers from our data 
set.

We supplement the {\it Swift} data with I, R, V, and B band fluxes and linear polarization
measured with the 1.83 m Perkins Telescope of Lowell Observatory in Flagstaff, AZ. The data acquisition, reduction, and analysis procedures are described by 
\citet{Jorstad2010}. The polarization is corrected for the interstellar value of $1.22\pm0.06\%$ in position angle $98^\circ\pm1^\circ$, as
determined from observations of three comparison stars in the field, which we assume
to be intrinsically unpolarized. 

We apply corrections for interstellar extinction with values from \citet{SF11} listed in 
the NASA Extragalactic Database (NED) for the I, R, V, B, and U filters. For the UV data, we adopt the interstellar extinction curve of \citet{Fitzpatrick1999} with $R_V=3.1$. The
values of extinction thus derived are $A_\lambda=$ 0.448, 0.645, 0.816, 1.079, 1.289, 
1.734, 2.412, and 2.132 for the I, R, V, B, U, UVW1, UVM2, and UVW2 bands, respectively.

\subsection{Light Curve at 37 GHz} \label{subsec:37GHz}

The 13.7-m radio telescope at Aalto University Mets\"ahovi Radio Observatory includes 
3C~120 in its regular monitoring program, measuring its millimeter-wave flux density 
multiple times per month during the period covered by the current study. The observations 
utilized a 1 GHz-band dual beam receiver with a central frequency of 36.8 GHz.
Flux densities are calibrated by observations of DR~21, with NGC7027, 3C~274, and 3C~84
serving as secondary calibrators. A detailed description of the
data reduction and analysis is given in \citet{Terasranta1998}.

\subsection{Millimeter-wave VLBA Imaging} \label{subsec:vlba}

Since early 2012, we have routinely (roughly once per month) observed 3C~120 with the
Very Long Baseline Array
(VLBA) under the VLBA-BU-BLAZAR monitoring project. Each 24-hour session includes 40-48
minutes of observations of 3C~120, split into 10-13 scans of 3-5 minute durations. After
correlation of the data from the ten antennas with the DiFX software correlator
at the Long Baseline Observatory site in 
Socorro, New Mexico, we calibrate the data with the Astronomical Image Processing System 
(AIPS) software provided by the National Radio Astronomy Observatory. We then use
\texttt{Difmap} \citep{Shepherd1997} to perform imaging and 
self-calibration through
an iterative procedure. This involves making a preliminary image that is then used in
the \texttt{Astronomical Image Processing Software} (AIPS) {\citep{vanMoorsel1996}} 
routine CALIB, which adjusts the left and right circularly polarized visibility
phases independently, guided by the image, before final imaging and self-calibration in 
Difmap. Calibration of the electric-vector position angle $\chi$ of the linear 
polarization is based on the multi-faceted approach of \citet{Jorstad2005}. \citet{Jorstad2017} list further details of the data analysis procedures.

In order to quantify the emission structure in the jet, at each epoch we use the
Difmap model fitting procedure to divide the source into a number of emission components,
each with a circular Gaussian intensity distribution. The starting model is the best-fit
model of the previous epoch. The model fit determines the position relative to the core, 
flux density, and full-width half-maximum (FWHM) angular size of each component.
\citet{Jorstad2017} discuss the uncertainties in the model-fit parameters.

\begin{figure}[ht!]
\plotone{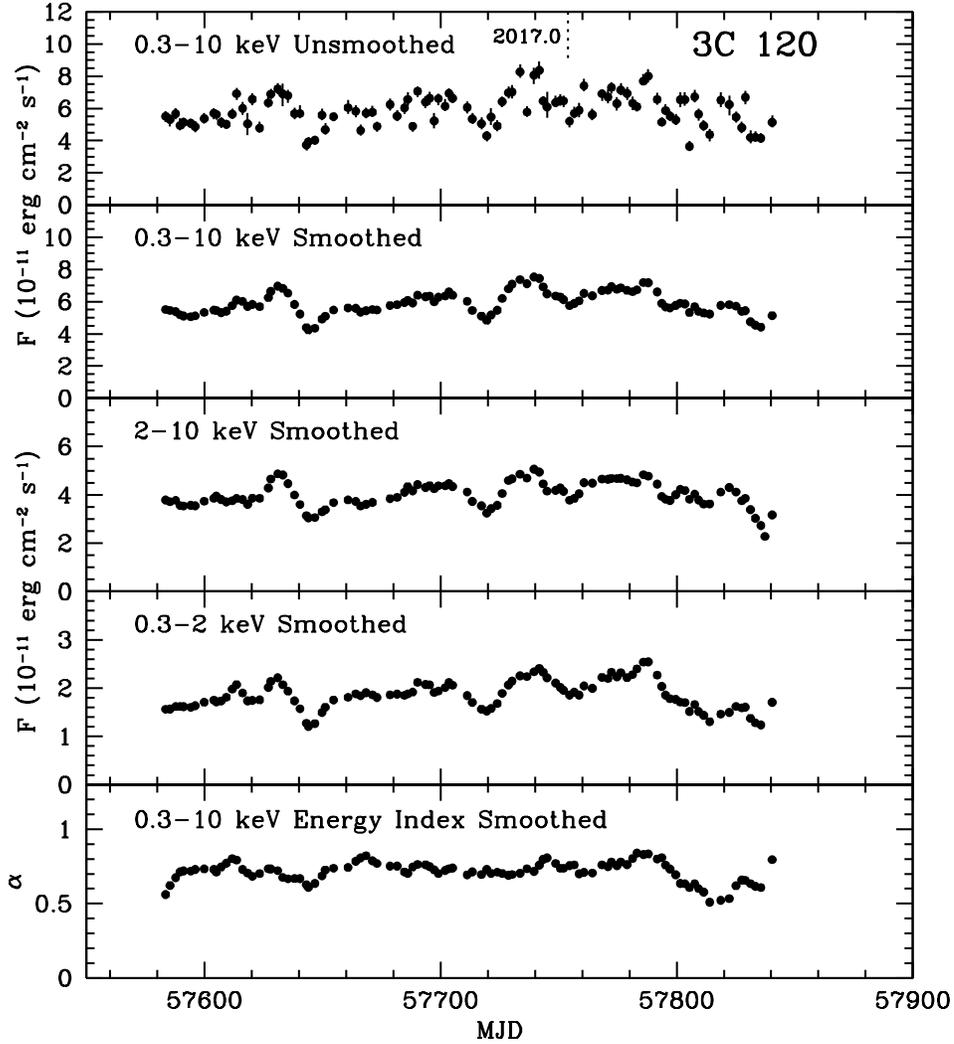}
\caption{Light curves of 3C~120 and (bottom panel) energy spectral index from
{\it Swift} XRT observations over the indicated X-ray energy ranges. Except for the top 
panel, the values plotted are smoothed over 5 consecutive observations (see text).
\label{fig1}}
\end{figure}

\section{Results} \label{sec:results}

\subsection{Light Curves} \label{subsec:lc}

Figure \ref{fig1} presents the {\it Swift} X-ray light curves, with the unsmoothed 0.3--10 keV data in the top panel, and smoothed 0.3--10 keV, 2--10 keV (hard X-ray), and
0.3--2 keV (soft X-ray) data in the next three panels. The bottom panel displays the
smoothed spectral index $\alpha$, where the flux density 
$F_\nu\propto\nu^{-\alpha}$. The smoothing is over five consecutive points, with
the flux at the epoch of the data point assigned 40\% weight, contiguous points given 20\%
weight, and points two epochs earlier or later given 10\% weight. (The first and last
data points are not smoothed, while the second and penultimate are smoothed over three
epochs.) The smoothed data
facilitate visual comparison with the UV and optical light curves, whose characteristic 
time-scales of variation are longer than that of the X-ray flux. However, in the
correlation analysis below, we use the unsmoothed values.

\begin{figure}[ht!]
\plotone{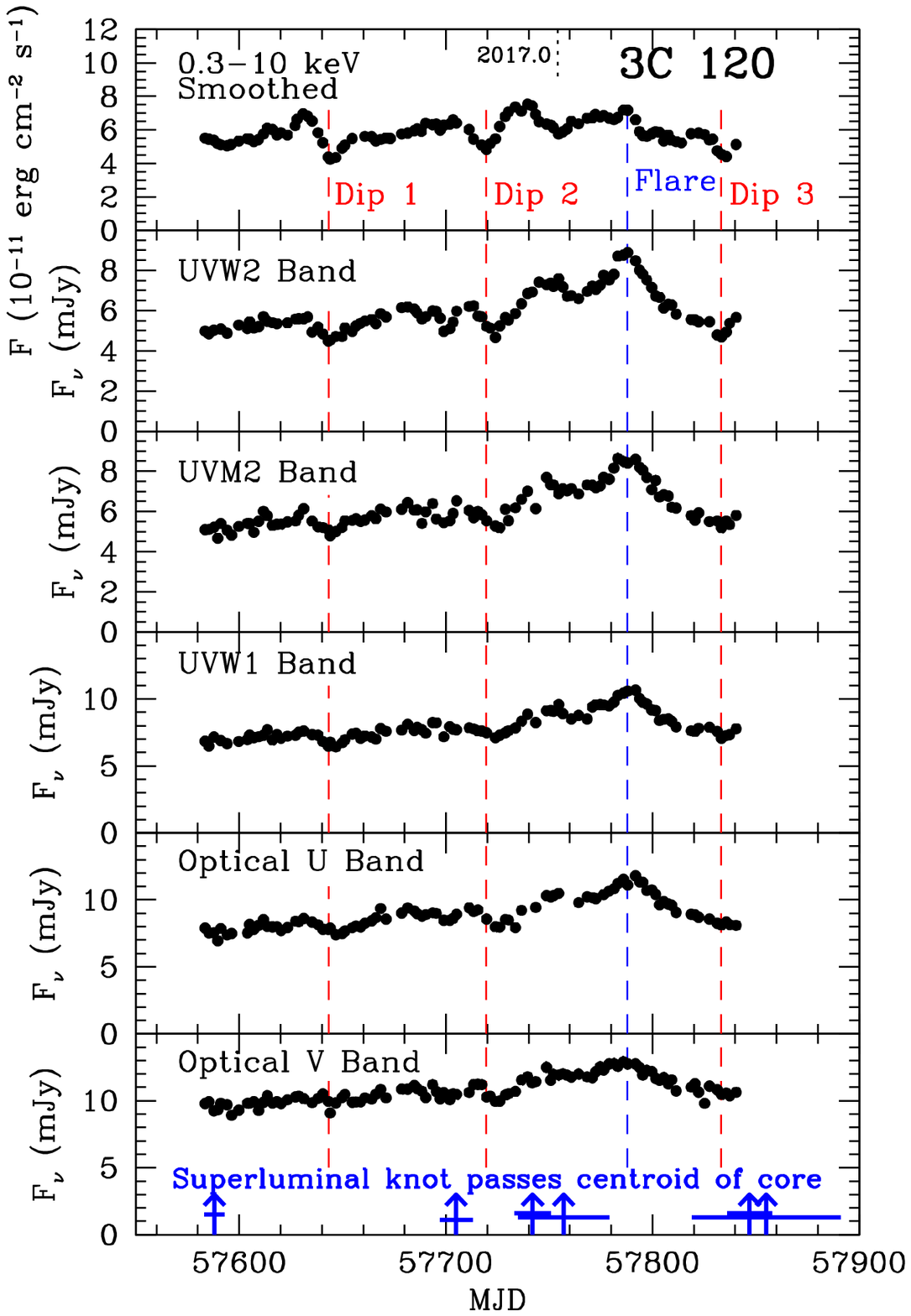}
\caption{Smoothed 0.3-10 keV X-ray and unsmoothed UV-optical light 
curves of 3C~120 from {\it Swift} observations. Three minima in the light curves are 
marked as ``dips'' with red vertical dashed lines, a flare is indicated by a
vertical dashed line, and epochs $t_0$ when a superluminal knot 
passed the ``core'' of the jet, as seen in 43 GHz VLBA images, are denoted by upward 
arrows in the bottom panel. The uncertainty in each epoch is denoted by a
horizontal line. \label{fig2}}
\end{figure}

\begin{figure}[ht!]
\plotone{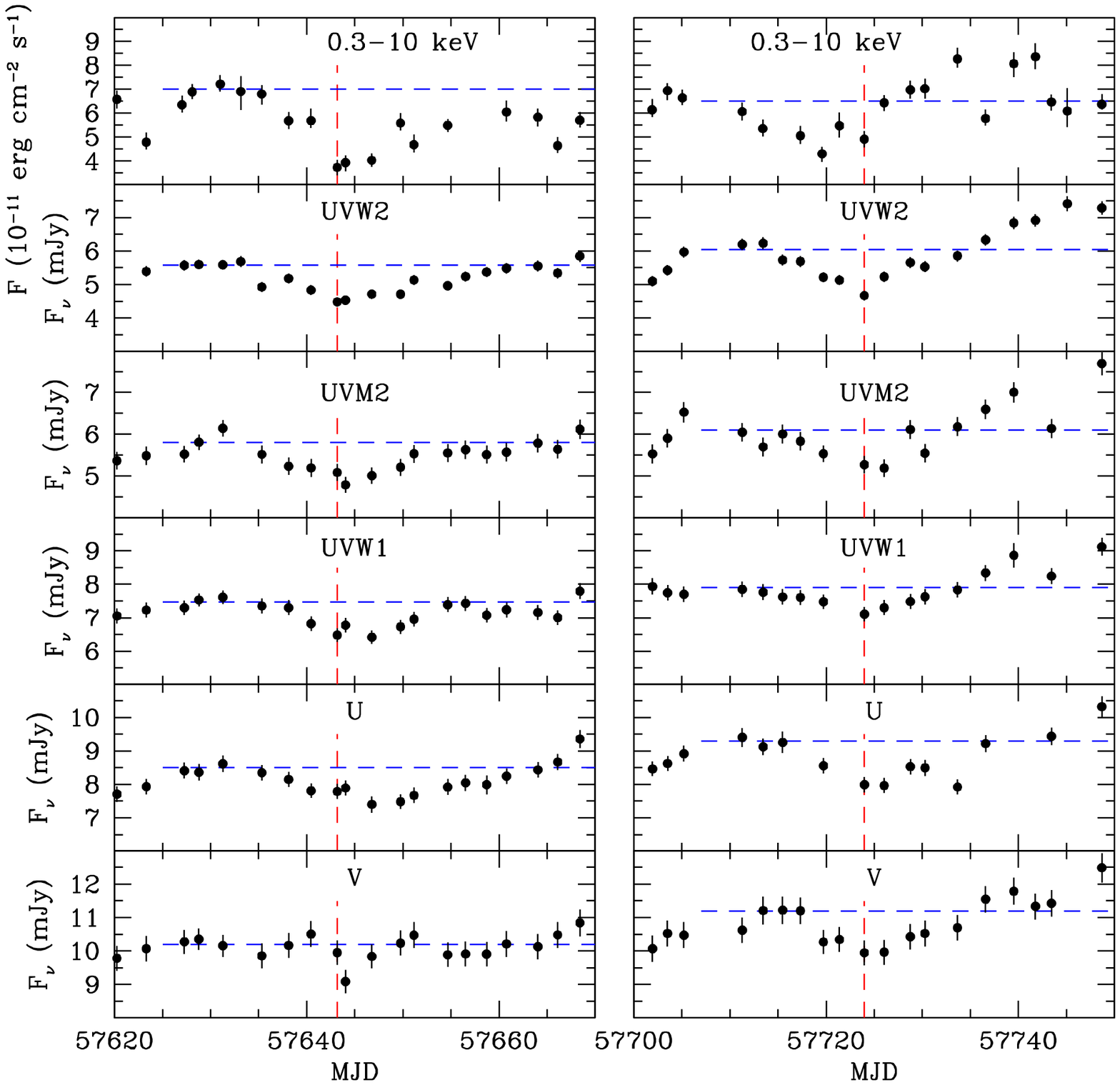}
\caption{Expanded view of the light curves (without
smoothing) over time spans that include dip 1 (left side) and dip 2 
(right side). The red vertical dashed lines mark the epochs of minimum UVW2 flux,
while the horizontal blue lines indicate the pre-dip flux level, obtained by
averaging the three data points immediately preceding the corresponding dip. \label{fig3}}
\end{figure}

Figure \ref{fig2} displays the UV and optical light curves of 3C~120, along with 
the smoothed 0.3--10 keV X-ray data. Minima in the light curves (``dips'') are denoted by
red vertical dashed lines and numbered for reference. In order to be classified as a dip,
the minimum in flux needs to be apparent at all bands. The local minimum near MJD 57760
is not included because the flux exceeds the mean flux in the UV-optical bands, and
the minimum appears to be part of a double-flare outburst rather than a dip as defined by
\citet{Chatterjee2009}. Figure \ref{fig3} expands the portions
of the light curves that include dips 1 and 2.

\begin{figure}[ht!]
\plotone{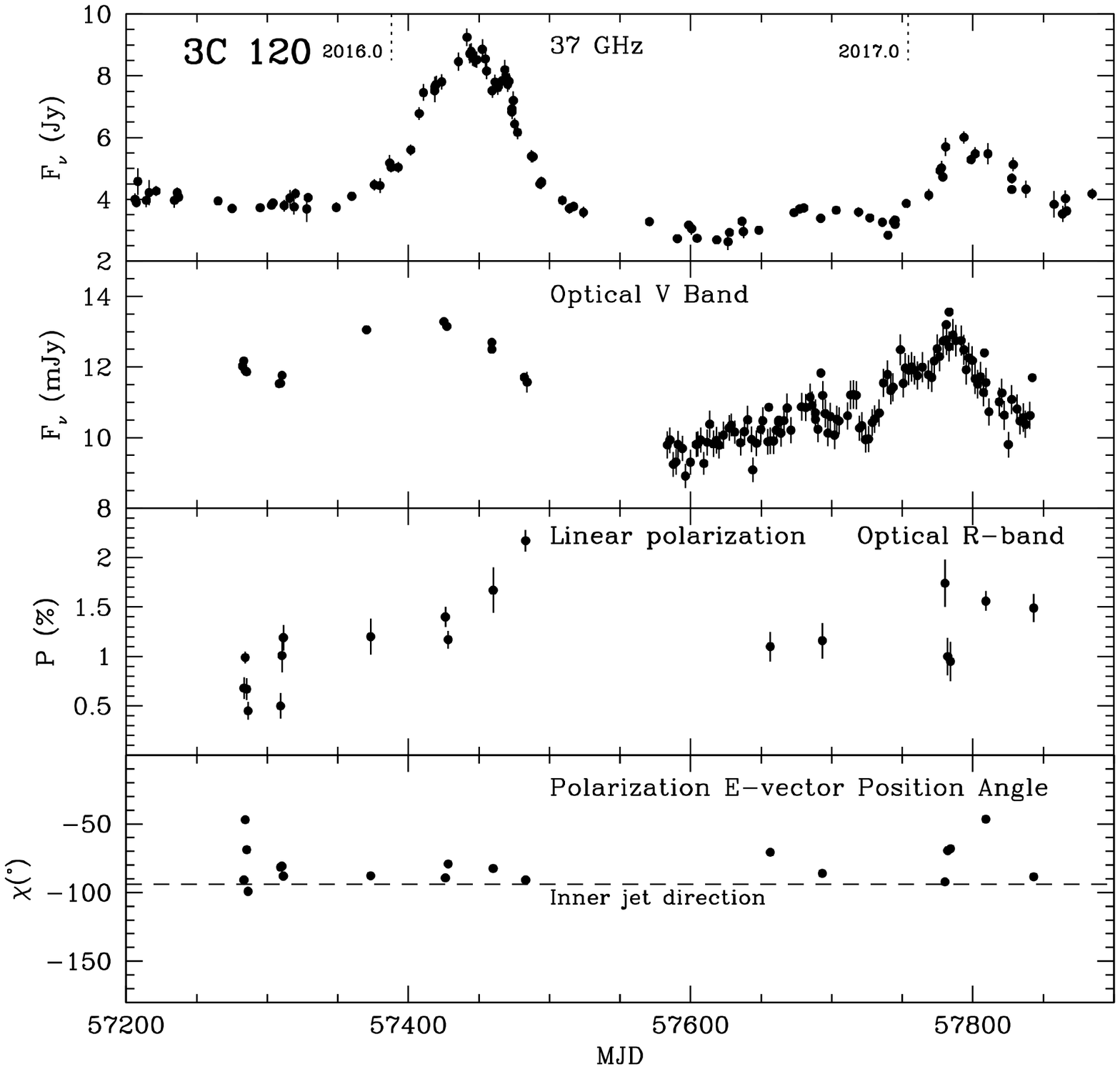}
\caption{{\it Top panel:} Light curve of 3C~120 at 37 GHz from measurements made 
at the Mets\"ahovi Radio Observatory. {\it Second panel from top:} Optical V-band light
curve; {\it bottom two panels:} degree and electric-vector position angle of optical 
R-band linear polarization. The direction of the innermost jet in 43 GHz VLBA images
is indicated in the bottom panel for comparison. \label{fig4}}
\end{figure}

The top panel of Figure \ref{fig4} presents the 37 GHz light curve of 3C~120 during our
{\it Swift} monitoring campaign, as well as during the previous year. The earlier data
include the highest-amplitude millimeter-wave outburst ever observed in 3C~120. In
\S\S\ref{subsec:radio} we compare the multi-waveband behavior during this event with
that of the strong outburst in 2017. For comparison,
the V-band optical light curve is also displayed, as are the R-band degree ($P$) and
electric-vector position angle ($\chi$) of linear polarization. There is a V-band
counterpart to each radio flare.

\begin{figure}[ht!]
\plottwo{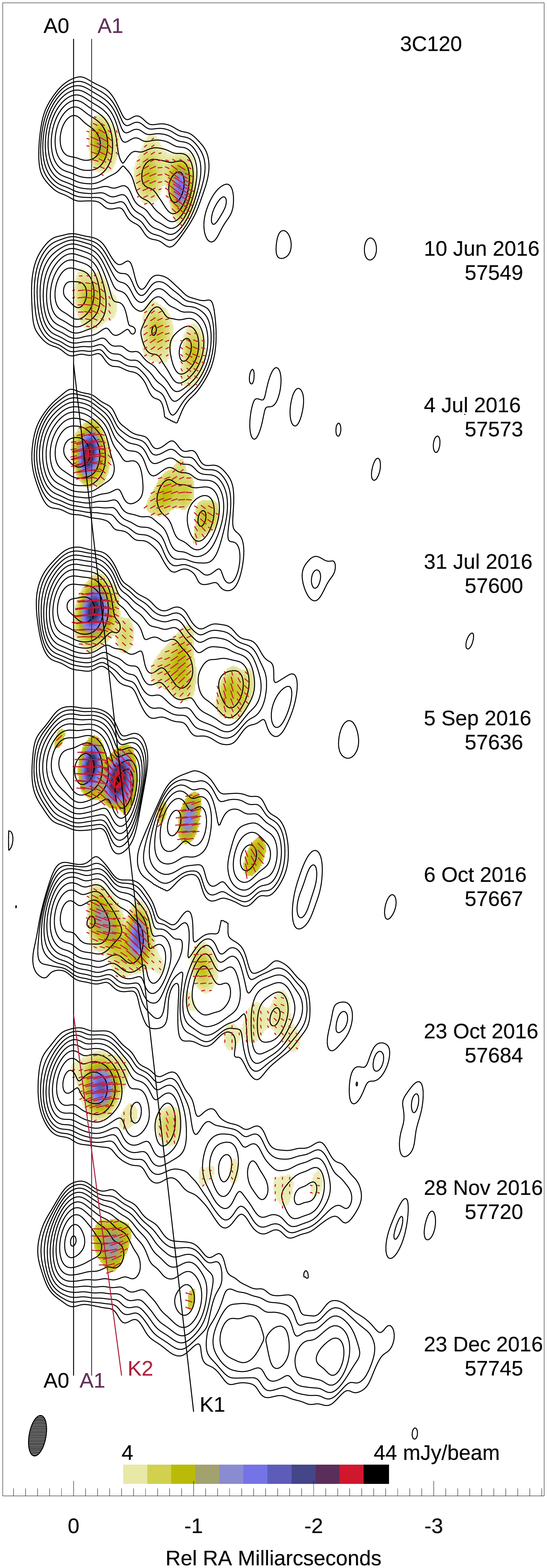}{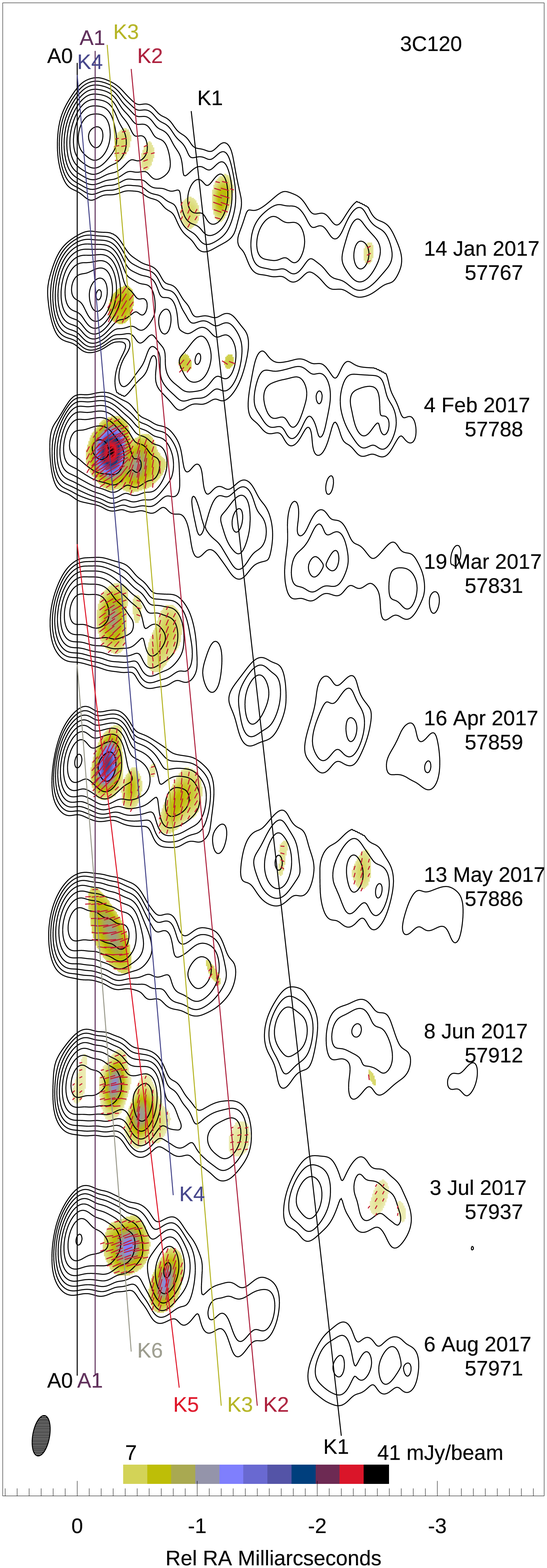}
\caption{VLBA images of 3C~120 at 43 GHz. Contours: total intensity in factors of 2
starting at 0.25\% of the global peak of (left panel) 1.17 Jy beam$^{-1}$ (attained on 2016 Oct 6) and (right panel) 2.41 Jy beam$^{-1}$ (attained on 2017 Feb 4), with
images at these two epochs including an extra contour at 95\%;
color: linearly polarized  intensity; red line segments: electric-vector
position angle of linear polarization. The ``core'' ($A0$), stationary feature $A1$,
and superluminal knots ($Kn$) are indicated.
The elliptical Gaussian restoring beam of FWHM dimensions $0.34\times0.14$ mas with
major axis along position angle $-10^\circ$ is displayed in the bottom left corner
of each panel.
\label{fig5}}
\end{figure}

\subsection{Changes in the Jet} \label{subsec:jet}

\begin{figure}[ht!]
\plotone{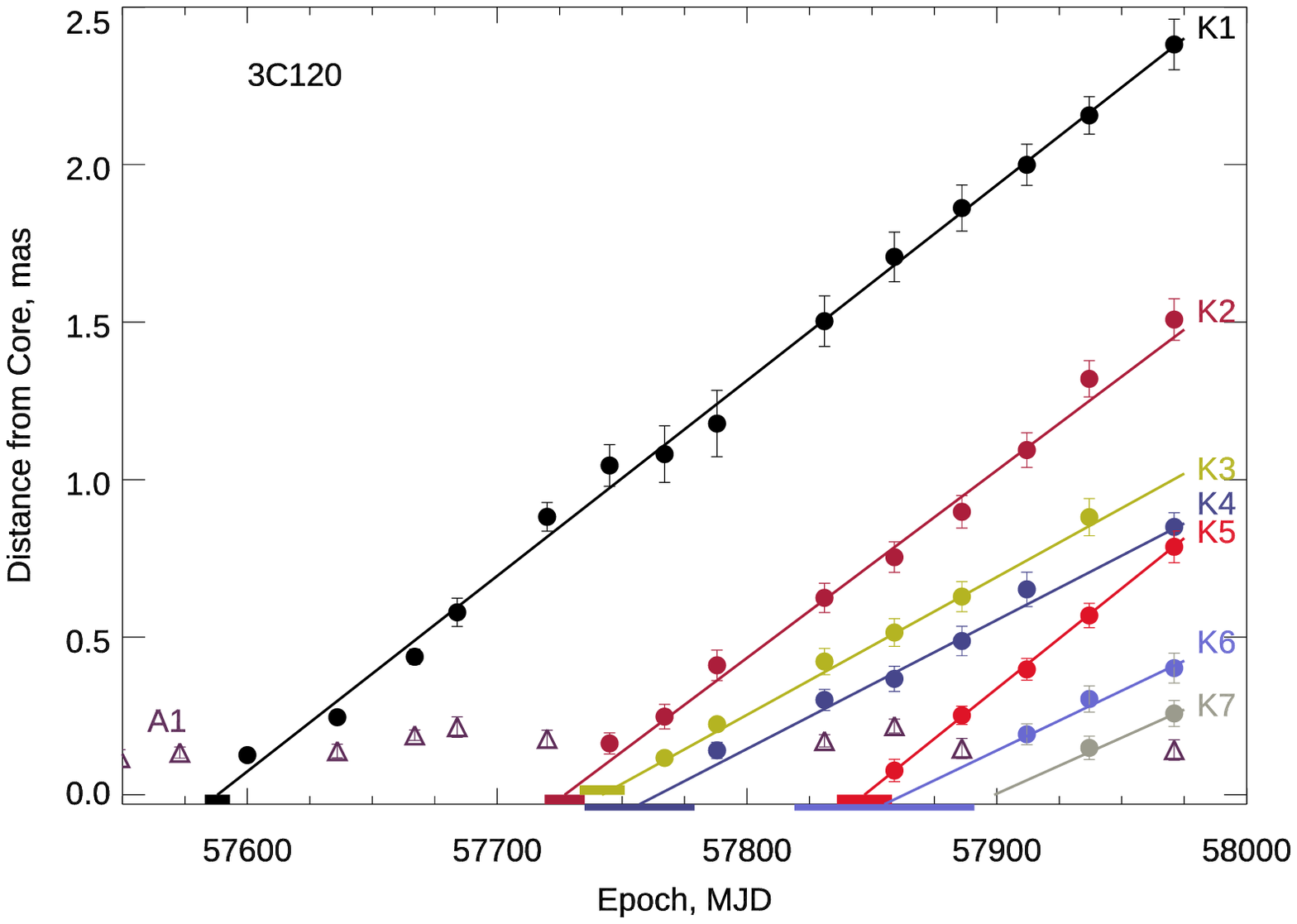}
\caption{Separations from the ``core'' of emission features identified in the VLBA images of 3C~120 at 43 GHz. Moving knots identified in Figure \ref{fig5} are coded in color and
labeled. The horizontal colored bars denote the range of times ($\pm$ uncertainties)
when the brightness centroid of the knot with the corresponding color was at the centroid of the core, as determined by a backward extrapolation of its motion.
\label{fig6}}
\end{figure}

Figure \ref{fig5} presents a time-sequence of 16 VLBA images at a frequency of 43 GHz from 
2016 June to 2017 August. The angular resolution of the longest baselines is $\sim 0.1$
mas along the direction of the jet. At a distance of 140 Mpc (for a Hubble
constant of 70 km s$^{-1}$ Mpc$^{-1}$), 0.1 mas corresponds to a length of 0.064 pc 
projected on the sky. For an angle $\theta$ between the jet and the line of sight 
\citep[$\sim 10^\circ$][]{Jorstad2017}, the deprojected conversion is 
0.1 mas = $0.36(\sin 10^\circ/\sin \theta)$ pc. The ``core'' is the bright feature
($A0$ in Fig.\ \ref{fig5}) at the eastern end of the
jet, and is presumed to be stationary. In compact extragalactic radio sources, the 
properties of the parsec-scale core are consistent with those expected from a standing 
conical shock in the jet \citep{Cawthorne2006,Cawthorne2013,Marscher2014,Marscher2017}. 
The images reveal a second quasi-stationary emission feature, $A1$, located 0.14 mas
(0.51 pc for $\theta= 10^\circ$) downstream of the core. Figure \ref{fig6} plots the 
separation of each knot from the core as a function of time (as 
determined by model fitting; see above), with a straight-line fit to 
the apparent motion. Extrapolation of the fit to zero separation yields the ``ejection'' 
time $t_0$ when the brightness centroid of each moving knot coincided with that of the 
core. Table \ref{motions} lists the apparent motions and values of $t_0$ for knots $K1$
to $K6$; there are insufficient data in Figure \ref{fig6} to determine the motion of $K7$.

\begin{deluxetable}{lllll}
\singlespace
\tablecolumns{5}
\tablecaption{\small\bf Apparent Motions of Moving Knots \label{motions}}
\tabletypesize{\footnotesize}
\tablehead{
\colhead{Knot}&\colhead{$\mu$}&\colhead{$v_{\rm app}$}&\colhead{$t_0$}&\colhead{$t_0$}\\
\colhead{}&\colhead{mas yr$^{-1}$}&\colhead{$c$}&\colhead{MJD}&\colhead{decimal yr}
}
\startdata
$K1$&$2.27\pm0.03$& $5.06\pm0.07$&$57588\pm5$&$2016.549\pm0.015$\\
$K2$&$1.97\pm0.05$& $4.84\pm0.11$&$57705\pm8$&$2016.830\pm0.021$\\
$K3$&$1.60\pm0.09$& $3.56\pm0.20$&$57742\pm9$&$2016.97\pm0.03$\\
$K4$&$1.44\pm0.13$& $3.17\pm0.29$&$57757\pm22$&$2017.01\pm0.006$\\
$K5$&$2.32\pm0.15$& $5.17\pm0.33$&$57847\pm11$&$2017.26\pm0.03$\\
$K6$&$1.29\pm0.29$& $2.88\pm0.64$&$57855\pm36$&$2017.28\pm0.10$\\
\enddata
\end{deluxetable}

\begin{figure}[ht!]
\plotone{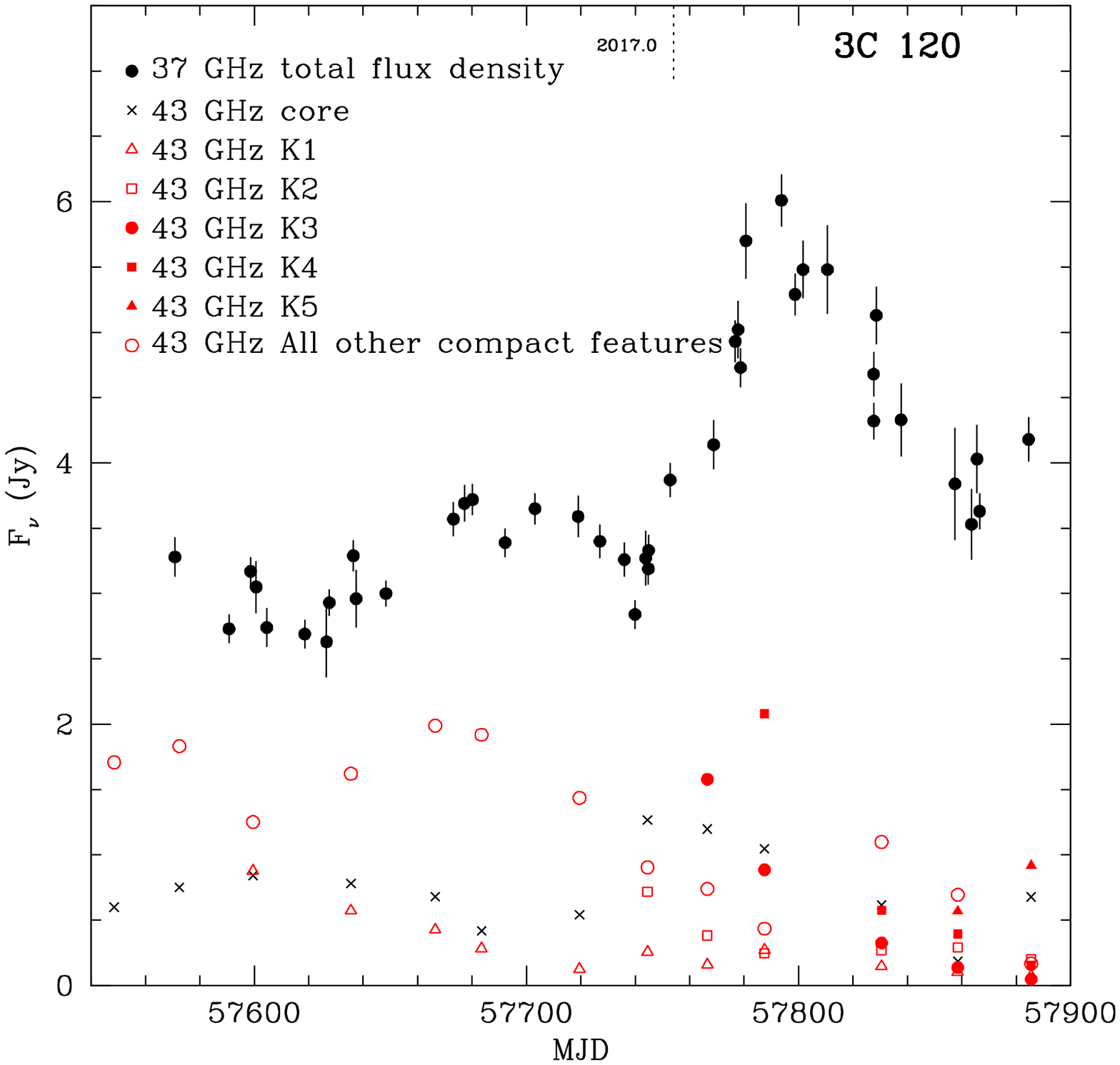}
\caption{Light curve of the entire source at 37 GHz along with the light curves at 43 GHz
of individual emission components identified in the VLBA images. Flux densities of the
components are derived from model fitting. At a given epoch, the sum of the individual flux densities may be less than that of the entire source because of errors in the VLBA
flux calibration and/or model fitting, as well as emission from low-intensity features not 
included in the model fitting.
\label{fig7}}
\end{figure}

Figure \ref{fig7} displays the 37 GHz flux density of the entire source versus time,
along with the flux densities of the brightest individual emission features identified
on the 43 GHz VLBA images. It is apparent that the maximum in flux density at MJD 57795 is mainly the result of increases in flux density of superluminal knots $K3$ and $K4$ when
they were 0.17 and 0.25 mas, respectively, downstream of the core (see Fig.\ \ref{fig6}).

\newpage
\subsection{Correlation Analysis} \label{subsec:cor}

In order to relate the flux variations across the X-ray, UV, and optical bands, we use
the z-transformed discrete correlation function maximum likelihood methods
ZDCF \citep{Alexander1997} and PLIKE \citep{Alexander2013},
which have been shown to determine correlations of 
unevenly sampled data effectively. In order to assess the statistical significance of the
correlations, we create 3000 artificial light curves (ALCs) at each waveband by using the 
actual data points. As in \citet{Williamson2014,Williamson2016}, active periods are
identified and randomly assigned start dates from the list of actual dates of
observation. After this procedure, which preserves the structure of flares present
in the light curves, the remaining observed fluxes are randomly assigned to vacant actual 
observation dates. The ALCs of the two wavebands being correlated are then randomly
paired and processed with the ZDCF routine.
A ZDCF value is considered to be statistically significant at the
95\% level if its magnitude exceeds that of 95\% of the artificial values.
The flux redistribution/random subset selection method \citep{Peterson1998} was also
employed as a check on the ZDCF results. We found complete consistency between the
two methods.

\begin{figure}[ht!]
\plotone{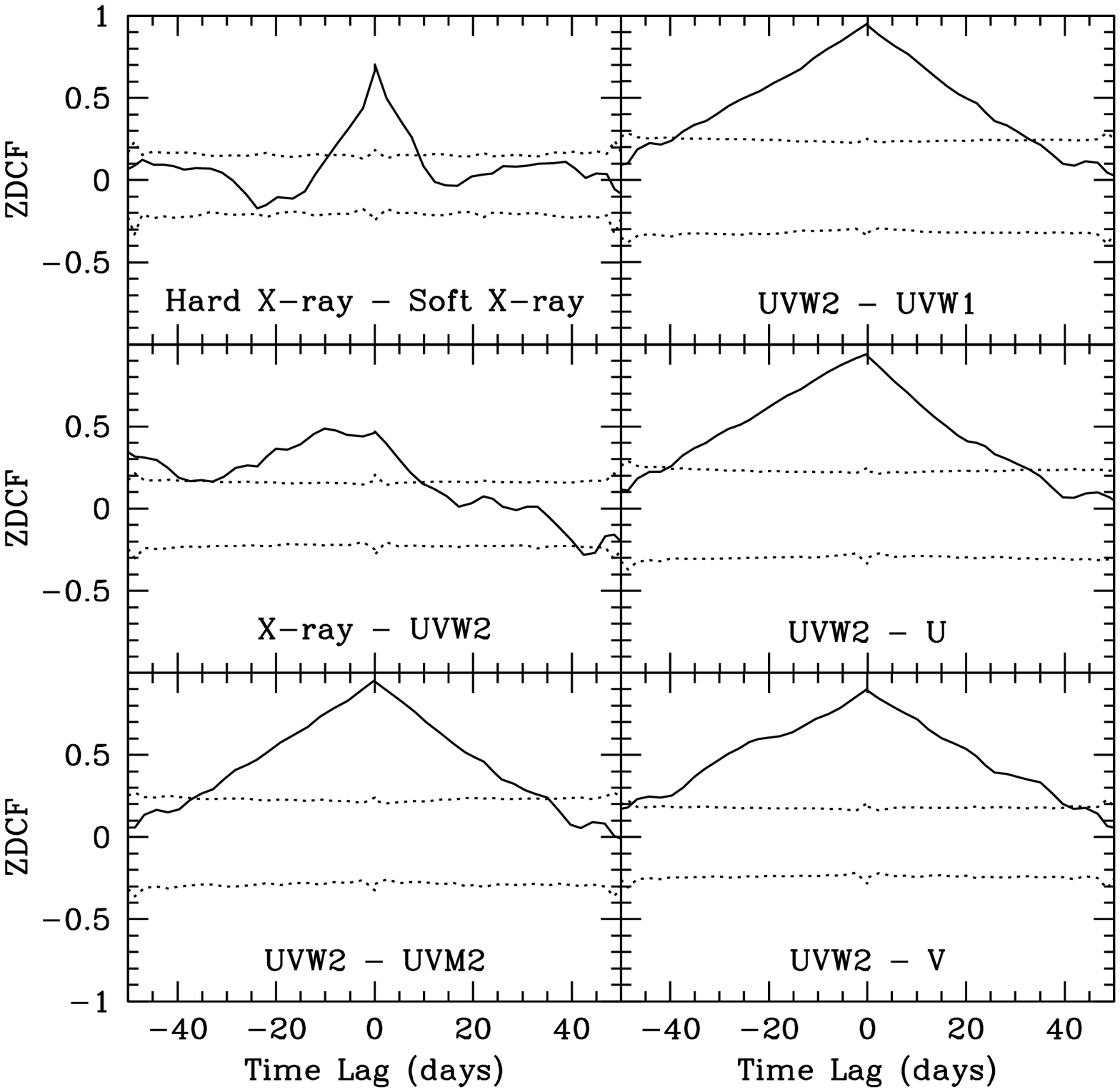}
\caption{ZDCF cross-correlations (normalized) of selected unsmoothed light curves as 
indicated in each panel. The hard X-ray photon energy range is 2-10 keV, while the soft 
X-ray range is 0.3-2 keV. The correlation functions are smoothed over
three lag times, with 50\% weight for the central time
and 25\% weight for each of the preceding and succeeding times. Negative lags
correspond to the first waveband listed leading the variations. Dotted curves delimit the 
95\% confidence level. See text for details. \label{fig8}}
\end{figure}

\begin{figure}[ht!]
\plotone{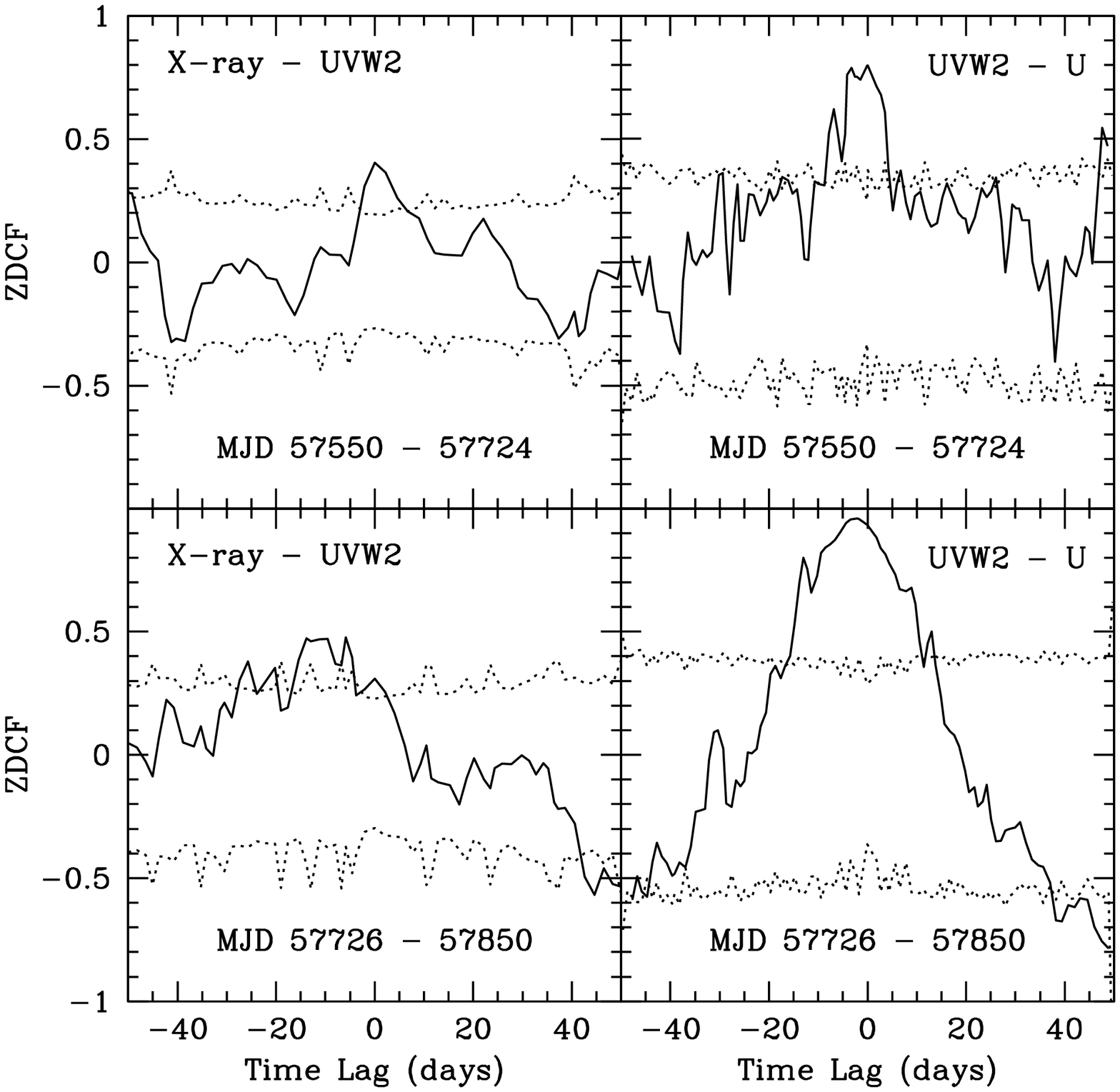}
\vskip 0.5cm \noindent
\caption{Cross-correlation function of (left) X-ray/UVW2 and (right) UVW2/U variations 
over time spans (top) MJD=57583-57724 and (bottom)  MJD 57724-57840.
Negative lags correspond to the first waveband listed leading the variations.
\label{fig9}}
\end{figure}

Figure \ref{fig8} presents the cross-correlations of the Swift data. The hard (2-10 keV)
and soft (0.3-2 keV) X-ray variations are well correlated, as are the UV and optical 
bands. The correlation between the X-ray and UV variations, however, is
more complex. As seen in the middle left panel in Figure \ref{fig8}, the X-ray/UVW2 correlation is moderately strong (peak ZDCF of 0.5) and flat from $-12$ days (X-ray 
leading) to zero. A non-zero lag over a $\sim 40$ day time span is apparent in the light 
curves exhibited in Figure \ref{fig2} starting with the event designated ``dip 2.'' In 
order to determine whether this change in lag is statistically significant, we perform the
correlation analysis separately on two halves of the light curves: from MJD 57550 to 57724 
and from 57726 to 57850. The result is displayed in Figure \ref{fig9}. The ZDCF for the
first time range peaks sharply at zero lag, with a somewhat lower maximum value of 
0.41 (at least partly caused by the lower flux level, which increases
the fraction of the flux contained in non-variable --- or slowly varying --- components 
such as starlight from the host galaxy).
In contrast, there is a maximum of 0.48 at $-6$ days and from $-10$ to $-14$ days
for the second time interval, with less pronounced peaks of 0.31, 0.36, and 0.38 at lags
of zero, $-20$, and $-26$ days, respectively. All of these local maxima of the ZDCF are
significant beyond the 95\% level. The difference in the correlations between the first
and second half of our monitoring period implies that the multi-waveband behavior
was more complex during the period of the double flare than during the previous, more
quiescent period.

We determine the cross-wavelength time lags by calculating the centroids of the 
cross-correlations, calculated over the area above 80\% of the maximum ZDCF value. We 
obtain the following lags, with negative lags corresponding to variations at the first 
waveband listed leading those at the second waveband: hard/soft X-ray: $-0.25\pm0.68$ 
days; X-ray(0.3-10 keV)/UVW2: $-6.2\pm2.4$ days; UVW2/UVM2: $-0.87\pm1.0$ days; UVW2/UVW1:
$-1.2\pm1.0$ days; UVW2/U: $-3.3\pm1.0$ days; and UVW2/V: $-1.4\pm1.6$ days. For the
correlations during two separate time periods displayed in Figure \ref{fig9}, we find
lags of: X-ray (0.3-10 keV)/UVW2: $+0.1\pm0.5$ days and UVW2/U: $-1\pm1$ days over
MJD 57550-57724, and X-ray (0.3-10 keV)/UVW2: $-10\pm3$ days and UVW2/U: $3\pm2$ days for
MJD 57726-57850. The lags between the X-ray and both UVW2 and U variations are 
consistent with those obtained by \citet{Buisson2017}, whose uncertainties are higher 
owing to sparser time coverage.

\subsection{Linear Polarization} \label{subsec:pol}

The optical linear polarization measurements are plotted in the bottom two panels of
Figure \ref{fig4}. Values range from $\sim 1$ to 2.2\%, with uncertainties of
$\pm0.1$-0.2\%. The electric-vector position angle $\chi$ lies within $\sim 45^\circ$ of 
the jet axis at all of the epochs. The VLBA images
presented in Figure \ref{fig5} indicate the polarized intensity and EVPAs at various
locations in the parsec-scale jet. As has been the case at most earlier epochs
\citep{Gomez2001,Gomez2011}, polarization is not detected in feature $A0$ (the 
``core'') at any of the epochs. This could result from either Faraday depolarization or a 
highly disordered magnetic field. Stationary feature $A1$ is significantly polarized only 
at epochs when a superluminal knot passes through it. The value of $\chi$ when this occurs
is usually $\sim 90^\circ$, parallel to the inner jet direction and
within $\sim 45^\circ$ of the optical EVPA.

\section{Discussion} \label{sec:discussion}

\subsection{Modeling the UV-Optical Emission} \label{subsec:mod}

The X-ray, UV, and optical light curves displayed in Figures \ref{fig1} and \ref{fig2} are
characterized by major dips and flares, as well as apparently random, lower-amplitude 
fluctuations. (Three dips and the most pronounced flare at UV-optical wavelengths are 
marked in Fig.\ \ref{fig2}.) This behavior is similar to that reported in 2002--08 by 
\citet{Chatterjee2009}, although the X-ray and R-band quiescent flux levels are
$\sim 20\%$ and $\sim 100\%$ higher, respectively, in 2016-17 relative to the average
levels in 2002-08. During the relatively quiescent period of MJD 57683-57717,
the error-corrected rms variability \citep[see][]{Buisson2017} after removal of a 
linear trend in V band: $0.81\pm0.30$ mJy ($14\pm5.2\%$), U band: $0.56\pm0.20$ mJy 
($9.7\pm3.5\%$), UVW1 band: $0.42\pm0.16$ mJy ($7.4\pm2.8\%$), UVM2 band:
$0.41\pm0.16$ mJy ($5.7\pm2.2\%$), UVW2 band: $0.52\pm0.14$ mJy ($6.7\pm1.9\%$), and 
X-ray (0.3-10 keV): $(8.1\pm3.0)\times 10^{-12}$ erg cm$^{-2}$ s$^{-1}$ ($14\pm5.2\%$). A 
least-squares straight line with $\alpha_{\rm rms}=0.48\pm0.22$ passes within $0.7\sigma$ 
of all of the UV-optical rms-variability data points, although the U, UVW1, and UVM2
measurements all lie below the line and the UVW2 value lies above. \citet{Buisson2017} 
measured a quite different slope for 3C~120, $\alpha_{\rm rms}=-0.21\pm0.10$
in 2011-12, similar to their values for radio-quiet Seyfert galaxies. This implies
a more significant steep-slope, variable UV-optical component in 3C~120 in 2016-17.

\begin{figure}[ht!]
\plotone{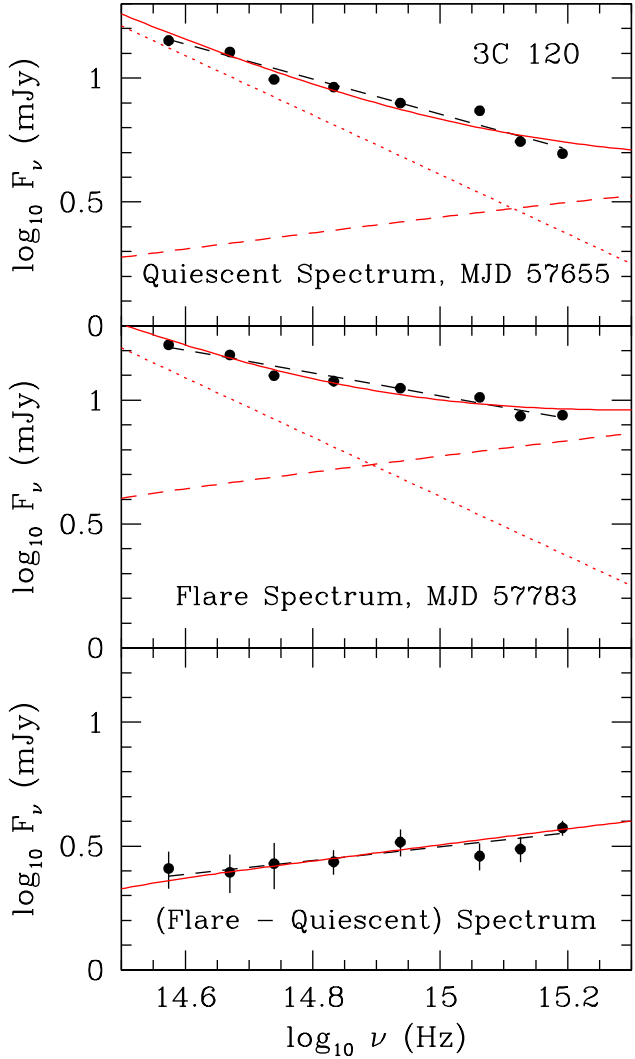}
\vskip 0.5cm \noindent
\caption{UV-optical spectrum during quiescence (MJD 57655, top panel), the peak of the 
double flare at MJD 57783 (middle panel), and the difference between the two
(bottom panel). The black dashed lines correspond to the 
best-fit single power laws, with spectral indices of $\alpha =$ $0.71\pm0.05$,
$0.46\pm0.03$, and $-0.28\pm0.07$ for the top, middle, and bottom panels, respectively,
where the flux density $F_\nu \propto \nu^{-\alpha}$. The red dotted line indicates
the spectrum of a model synchrotron component with $\alpha = 1.2$. The red dashed line
corresponds to the computed spectrum of a standard thin accretion disk (AD) model with
inner radius $r_{\rm in} = 1.3\times10^{13}$ cm and outer radius $r_{\rm out} = 1.3\times10^{16}$ cm, with a temperature $T(r) = T_{\rm in} r^{-3/4}$. (An identical
spectrum would result from synchrotron emission from a nearly mono-energetic population
of electrons with energies exceeding $\sim 10$ GeV.) In the top panel,
$T_{\rm in} = 3.0\times10^5$ K, while in the middle panel, $T_{\rm in} = 4.0\times10^5$ K.
The solid red curve shows the sum of the synchrotron plus AD models.
\label{fig10}}
\end{figure}

Between dips 2 and 3, the X-ray, UV, and optical light curves are dominated by a double 
flare. The top panel of Figure \ref{fig10} displays the UV-optical spectrum on MJD 57655, 
a date during a relatively quiescent period when we obtained data from all optical and UV 
bands listed in \S\S\ref{subsec:ouv}. The middle panel of Figure \ref{fig10} presents the
spectrum at the peak of the second flare on MJD 57783. (We have not subtracted 
the flux of emission lines or starlight from the host galaxy.) The spectrum at both epochs 
can be fit by a simple power law that is flatter during the flare, when the spectral 
index $\alpha = 0.46\pm0.03$ compared with the quiescent value of $0.71\pm0.05$. The power 
laws are too steep and extend to frequencies too high to be explained by a superposition 
of blackbody emission from hot dust, starlight, and the AD, as found in several quasars 
\citep{Kishimoto2008}. Instead, the spectra are consistent with synchrotron radiation,
at least at the longer optical wavelengths. In this case, the flattening of the UV-optical 
spectrum during the flare would correspond to an increase in efficiency of acceleration of 
electrons toward higher energies. This interpretation is consistent with the optical 
linear polarization of 1-2\% measured during the {\it Swift} monitoring
(see Fig.\ \ref{fig4}), as is discussed in \S\S\ref{subsec:radio}.

The zero lag between the X-ray and UV-optical variations during the quiescent period
(see \S\S\ref{subsec:cor} and Figs.\ \ref{fig8}-\ref{fig9}) requires the UV emission region to lie within 0.5
lt-days (the uncertainty in the peak lag value) of the X-ray source, and could be 
coincident with it. One possibility is that the X-ray and UV-optical emission both
originate in the base of the jet, as has 
been proposed for X-ray binaries by \citet{Markoff2005}. However, models for relativistic
jet formation that rely on strong helical magnetic fields within $\sim 1000 R_g$ of the
base \citep[e.g.,][]{BZ1977,BP1982,VK2004,MN2007,TNM2011} predict a high degree of linear 
polarization unless the jet points within $\sim 2^\circ$ of our line of sight 
\citep{Lyut05}, contrary to observations \citep[see above;][]{Jorstad2017}. 

An interpretation that explains various aspects of the data incorporates two main 
UV-optical emission components: an inverted-spectrum (IS) source
plus a steep-spectrum synchrotron source. In order to assess whether
the AD could be the IS source, we integrate the Planck function over radius to
produce a thin-disk spectrum, and add a power-law nonthermal spectrum
$F_\nu \propto \nu^{-\alpha}$ for both the
quiescent and flaring states. The best-fit model, which is compared to the observed 
spectra in Figure \ref{fig10}, combines a synchrotron component with $\alpha=1.2$
with an AD of inner radius $r_{\rm in} = 1.3\times10^{13}$ cm ($1.6 R_g$, where
$R_g = GM/c^2$ is the gravitational radius for a black hole of mass $M$), outer radius
$r_{\rm out} \geq 1.3\times10^{16}$ cm (5.0 lt-days), and temperature as a function of 
distance $r$ from the black hole $T(r) = T_{\rm in}(r/r_{\rm in})^{-3/4}$. In the model,
$T_{\rm in} = 3.0\times 10^5$ K at the quiescent epoch MJD 57655. The thermal component 
has an inverted UV-optical spectrum with the standard AD spectral index of 
$\alpha=-1/3$. In order to reproduce the flattening of the spectrum during the flare, we 
increase the flux of only the AD component by raising $T_{\rm in}$ by 33\% to $4.0\times 
10^5$ K at the peak of the flare while keeping the synchrotron component constant.
A smaller range of AD radii would result in curvature of the spectrum
well beyond the observational constraint. A larger value of $r_{\rm in}$ would increase
the flux by increasing the area of the disk at a given temperature, and would thus
require a small area filling factor, $(1.3\times10^{13}~{\rm cm}/r_{\rm in})^2$, to match the 
observed flux level. The flare minus quiescent spectrum, displayed in the bottom panel of 
Figure \ref{fig10}, rises with frequency, and can be fit by a power law 
with spectral index $\alpha = -0.28\pm0.07$, consistent with the theoretical value
of $-1/3$. (Note that this difference spectrum subtracts any steady emission
component such as starlight from the host galaxy.)

The value of $T_{\rm in}$ agrees roughly with the expectations of an AD 
model. When irradiated by a small (compared to the AD dimensions) X-ray source with 
luminosity $L_x$ located at a height $h_x$ above the center of the disk, the model gives
\begin{equation}
T(r) \approx [{{3GM\dot{M}}\over{8\pi\sigma r^3}} + {{(1-a)L_xh_x}\over{4\pi\sigma r_x^3}}]^{1/4},
\end{equation}
\citep{Cackett2007,Lira2011}, where $r_x$ is the distance between the X-ray source and 
disk at radius $r$ (assumed to be greater than $h_x$), $M$ is the
mass of the black hole and $\dot{M}$ is its accretion rate, $L_x$ is the luminosity of the 
X-ray source, $a$ is the albedo of the disk, and $\sigma$ is the Stefan-Boltzmann 
constant. In the quiescent state of 3C~120, $L_x \approx 2\times 10^{44}$ erg s$^{-1}$,
where we have assumed a spectral index of 0.87 up to a cutoff energy of $\sim 100$ keV
measured by NuSTAR at an earlier epoch \citep{Rani2018}, which is similar to the spectral
indices derived from our {\it Swift} observations (see Fig.\ \ref{fig1}, bottom panel).
For the albedo, which is poorly known, we adopt $a\sim0.2$.
For the accretion rate, we estimate $\dot{M}\sim 0.013 M_\odot$ yr$^{-1}$ by dividing the 
quiescent value of the far-UV luminosity, $\sim 3\times 10^{44}$ erg s$^{-1}$, by 
$c^2$ and by an assumed efficiency of 40\% of conversion of accretion power into 
radiation for a rapidly spinning black hole \citep{Shakura1973}. The resulting value of 
$T_{\rm in}$ is $\sim 3\times 10^5$ K, in agreement with the value from our spectral model 
during quiescence. However, the increase in temperature of the AD by 33\% in the
spectral model of the flare exceeds the maximum possible increase (when the second term
of eq.\ 1 dominates) of 10\% resulting from an increase in X-ray luminosity by the
observed flare:quiescent ratio of $\sim 1.5$. A more complex, or different, model is 
therefore needed to explain the high-amplitude variations in UV-optical flux.

The level of fluctuations in flux during the quiescent period (see above) agrees with 
the two-component model. There is a minimum in the fractional flux variability in the UVW1 
band, which is near the wavelength where the IS and synchrotron source contribute nearly 
the same fraction of the total flux density. If the synchrotron and IS fluctuations are
unrelated, then they will often partially cancel each other near that wavelength, whereas
at shorter and longer wavelengths the fluctuations are dominated by a single component
and therefore more coherent. In the case of fluctuations corresponding to random noise, 
the sum of flux variations from two components with equal variability amplitude is
expected to have an rms $\sim 1/\sqrt{2}$ times lower than that of a single component. The 
reduction in rms from V band to UVM2 and increase from UVM2 to UVW2 
during the relatively quiescent period (see above) agree with this expectation.

Qualitative support for the proposal that the AD is the IS source can be found in time lags of the multi-wavelength variability 
of the flux density. As seen in the top-right panel of Figure \ref{fig9}, during the relatively quiescent period of MJD 57550-57724, the UVW2 variations lead the U-band
variations by $-1\pm0.5$ days, with peaks in the ZDCF at both 0 and $-4$ days. During the 
flaring period (bottom-right panel of the figure), the ZDCF is broader, with a peak at $-3\pm1$ days. 
This delay toward longer wavelengths is consistent with the AD model if variations in flux 
propagate from inner to outer radii at speeds $<c$. However, the longer UVW2/U
delay relative to the UVW2/V lag is naturally explained as scattering via Balmer
continuum emission by clouds a few lt-days from the inner AD \citep{Cackett2018}.
Furthermore, as is discussed in \S\S\ref{subsec:timescale} below, cross-wavelength time 
lags $\gtrsim0.5$ days are much longer than predicted by the basic AD model, which 
therefore needs additional features in order to explain all of the data.

\begin{figure}[ht!]
\plotone{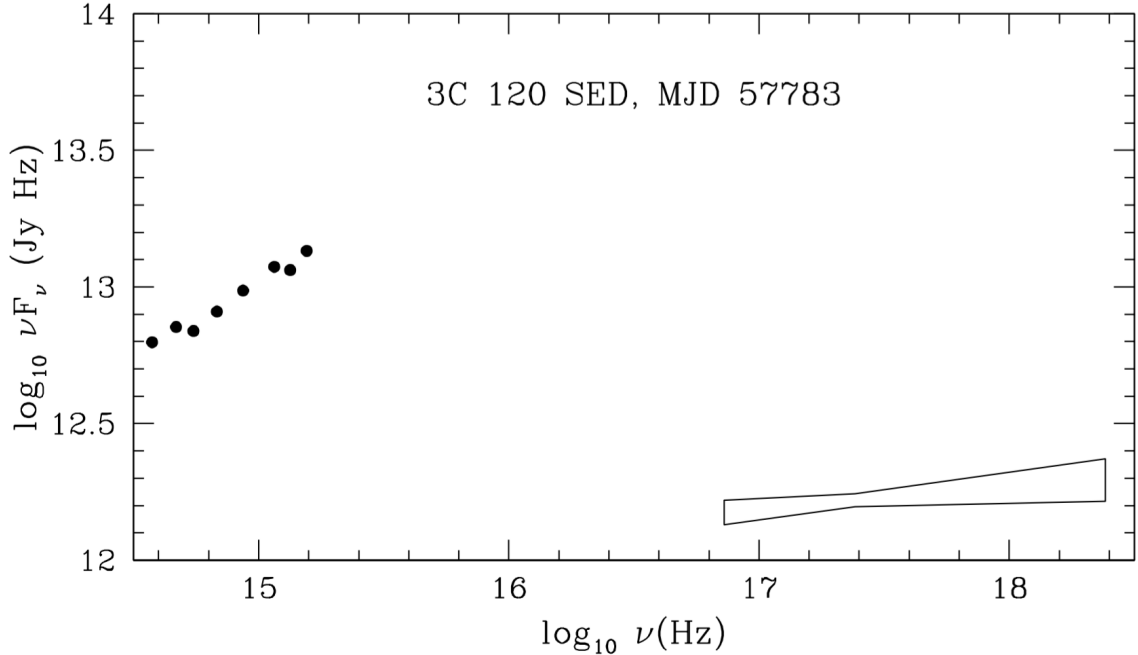}
\vskip 0.5cm \noindent
\caption{Spectral energy distribution at the peak of one of the flares; no 
soft X-ray excess is apparent. \label{fig11}}
\end{figure}

The spectral index of the flare-minus-quiescent spectrum is also similar to the value
of $-1/3$ for optically thin synchrotron radiation from electrons whose critical 
frequencies are all higher than $\sim 2\times 10^{15}$ Hz \citep[see][]{Pach1970}.
The minimum energy of such an electron population would need to exceed
$\sim10B^{-1/2}$ GeV, where $B$ is 
the magnetic field strength in Gauss. The maximum energy could not be greater than
$\sim 3$ times higher than the minimum. Otherwise, the soft X-ray synchrotron flux would 
exceed the observed value; cf.\ the spectral energy distribution (SED) during the flare, 
displayed in Figure \ref{fig11}. Such a sharply peaked SED is a possible outcome of
particle acceleration by magnetic reconnection \citep[see][]{Petro2016}. In fact,
models of very high-energy $\gamma$-ray emission from BL Lacertae objects --- which
are associated with FR~1 radio sources --- generally require minimum energies of
$\sim 10$ GeV \citep[e.g.,][]{Balokovic2016}. In \S\S\ref{subsec:radio} below, we find
that this alternative model for the IS component during the flaring period is consistent 
with a strong correlation found between the UV-optical and 37 GHz variations of 3C~120.  

\subsection{Dips in Flux} \label{subsec:dips}

As is discussed in \S\ref{sec:intro}, the inner-disk disruption scenario for 3C~120 proposed 
by \citet{Lohfink2013} predicts that, during dips in flux,
the UV flux should decrease a short time before the X-ray
flux. Since the inner disk is refilled from larger to smaller radii, the recovery of
the flux to its pre-dip value should proceed from longer to shorter wavelengths at a
propagation speed $v_{\rm refill}\ll c$.

As seen in Figure \ref{fig3}, during dip 1 the UVW2 
flux decreased by 15\% prior to the first significant decrease (by about the same 
percentage) in the X-ray flux one observation (2.2 days) later. However, a pre-dip X-ray 
flare, which is also apparent in the UVM2 data, took place immediately before the dip, 
compromising our ability to determine the start time of the X-ray dip. Compared with the 
UVW2 flux, the UVM2 and UVW1 fluxes reached minimum later, while the U-band dip was 
delayed by 4 days and its recovery was completed 
1-2 days later. The light curves at different
wavelengths therefore possess a similar overall pattern, as well as cross-wavelength 
discrepancies, indicative of overlapping---but not identical---emission regions. While
the earlier initial drop in UV relative to X-ray flux agrees with the \citet{Lohfink2013} 
picture, as does the later minimum at longer wavelengths, the recovery does not occur
earlier at longer wavelengths, contradicting the model.

A superluminal knot ($K2$; see Figs.\ \ref{fig5} and \ref{fig6}) passed through the
core in the 43 GHz VLBA images about 75 days after the start of dip 1. This conforms with
the range of time delays reported by \citet{Chatterjee2009}, who found a mean delay of
$68\pm 14$ days. At the apparent speed of $K2$, 2.0 mas yr$^{-1}$ = $4.3c$,
the knot traveled about 1.3 pc before reaching the centroid of the core if the angle 
between the velocity vector and line of sight is $\theta\sim10^\circ$. A shorter 
distance to the core was derived in \citet{Chatterjee2009} based on a wider viewing angle.

In contrast with dip 1, dip 2 (see Fig.\ \ref{fig3})
started with a decline in the X-ray flux, which reached its 
lowest level (34\% below the pre-dip flux) 4.4 days before the UVW2 minimum (23\% drop). 
As opposed to dip 1, the X-ray recovery led that at 
UV-optical wavelengths, although the immediate transition from the dip to a flare
complicates the interpretation if the flare occurred through another process at another
location. The early X-ray
start implies that the dip was initiated by changes in the corona rather than the AD.
This conclusion is compatible with the \citet{Chatterjee2009} scenario (see 
\S\ref{sec:intro}). However, the
UV-optical dips and recoveries were simultaneous to within the uncertainties. This,
alongside the $>4$ day time lag between the X-ray and UVW2 dips, is difficult to 
understand if the change in physical conditions in the AD propagates at a speed $\ll c$. 

As was the case for dip 2, dip 3 began with a decrease in X-ray flux. However, our 
observational campaign ended before the flux recovered completely. Hence, we cannot 
perform a meaningful analysis of the behavior of the emission over the entire event. 

A pair of new superluminal knots appeared after both dips 2 ($K3$ and $K4$) and 3
($K5$ and $K6$). It is possible that
these knots are examples of shock pairs, with both a forward ($K3$ and $K5$) and
reverse ($K4$ and $K6$) shock.
\citet{Casadio2015} have interpreted a previously observed closely-spaced pair of 
superluminal knots in 3C~120 
in this manner. Such a shock pair can arise from a disturbance caused by an increase in
the flow velocity, as expected in the proposal of \citet{Chatterjee2009}. This
scenario predicts that the first knot should be faster than the second for each pair, as
is the case (see Fig.\ \ref{fig6}). However, the delay between the starts of dips 2 and 3 
and the passage of leading knots $K3$ and $K5$
through the core was considerably shorter than for dip 1 and knot $K2$: $32\pm 6$ days and 
$20\pm 6$ days, respectively. This is compatible with higher actual speeds in these
knots compared with $K2$ only if the angle $\theta$ between the velocity vector and line 
of sight is much less than for $K2$, since the apparent speed of $K2$, $4.3c$, is greater 
than that of $K3$, $3.4c$. For example, an increase in the bulk Lorentz factor
$\Gamma$ from
4.5 for $K2$ to 5.3 for $K3$ while $\theta$ decreased from $11^\circ$ to $4^\circ$
would correspond to the observed decrease in the apparent speed and reduce the apparent
transit time to the core by 50\%. This would cause an increase in the Doppler beaming 
factor from $\sim 5$ to $\sim 9$, leading to knots $K3$ and $K4$ having 
much higher fluxes than their predecessors. The high-amplitude radio flare apparent in 
Figure \ref{fig4} is consistent with this scenario: Figure \ref{fig7} demonstrates that
the flare is the result of knots $K3$ and $K4$ appearing in the jet at 43 GHz with
fluxes that are extraordinarily high for superluminal knots in 3C~120.
This proposed change in $\theta$ is quite large and would need to occur over a
short time span. On the other hand, such a possibility might be indicated by the 
discrepant values from $\theta = 20^\circ$ to $3.6^\circ$ derived from previous VLBA 
observations \citep{Jorstad2005,Casadio2015,Jorstad2017}. The proposed change in actual
velocity of the flow corresponds to a relative velocity of $0.17c$, 0.29 times the
sound speed of an ultra-relativistic plasma. This is consistent with the formation of a
shock if the plasma contains enough non-relativistic protons to reduce its sound speed to
$<0.29c$.

Under the magnetic jet launching scenario, all X-ray/UV-optical dips could arise from 
occasional, probably randomly occurring, alignment of the magnetic field along the polar 
direction in the AD and corona \citep{Livio2003}. According to these authors, this would 
disable the magneto-rotational instability, disrupting the accretion flow while promoting 
faster flow in the polar direction into the corona and jet. 
In dip 1, the earlier UV flux decline suggests that the alignment occurred first in the
inner AD and then propagated (presumably at the magnetosonic speed) into the corona. In
dip 2, the data imply that the alignment occurred first in the innermost AD and corona
and propagated outward through the AD.

Random fluctuations in local conditions (e.g., density or
temperature of the Compton scattering electrons in a section of the corona), superposed
on the systematic trends (such as a short-term X-ray and UVM2 flare observed prior to dip 
1), could cause enough complexity in the light curves to lead us to infer an incorrect
start and end time of dips. Future observations of more such events would provide better 
statistics on the time delays across wavelengths.

\subsection{Time-scales of Variations and Cross-Wavelength Lags} \label{subsec:timescale}

The time lags predicted by the AD model proposed in \S\S\ref{subsec:mod} are all
$< 1$ day, since the inner radius is only 0.12 light-hours and the radius where the
blackbody spectrum peaks in V band is only 100 times larger, 0.5 lt-days. (For an
area filling factor $f$, this size should be multiplied by $f^{-1/2}$.) This
is consistent with the UV/optical lags over the period MJD 57550-57724 reported in
\S\S\ref{subsec:cor} and seen in the top panels of Figure \ref{fig9}, although there is
some longer lag present in U band that we address below. However, longer 
lags are present from MJD 57726 to 57850 when slower, higher-amplitude variations 
occurred. This included the flare that peaked near MJD 57783. It therefore appears that the shorter-term variations are essentially simultaneous at the different 
wavelengths to within the limits of our sampling, while longer-term flares tend 
to lag at longer wavelengths by up to $\sim10$ days. We present evidence in 
\S\S\ref{subsec:radio} that these flares occur in the parsec-scale jet rather than the AD.

Recent well-sampled, long-term {\it Swift} monitoring observations of the Seyfert galaxies 
NGC5548 \citep{McHardy2014,Edelson2015}, NGC4151 \citep{Edelson2017}, and NGC4593
\citep{Cackett2018,McHardy2018} have measured time lags between hard and soft X-ray (in
some cases), UV, and optical variations, with longer-wavelength
variations delayed. The ratios of the delays follow those expected from standard
thin AD models \citep{Shakura1973}, but the lags at individual frequencies
are a few times longer than the model would predict. 
The mass of the black hole in 3C~120 is within 30\% of those in NGC5548 and NGC4151,
\citep[although $\sim 7$ times that in NGC4593][]{Bentz2015,Edelson2017},
hence one might expect similar temporal behavior. It is therefore interesting that the
correlations over MJD 57550-57724 are consistent with the $< 0.5$-day time lags
predicted by the AD model, in contrast with the case of the Seyfert galaxies. There is
one exception, however: the U-band variations lag those in the UVW2 band.
\citet{Cackett2018} and \citet{McHardy2018} propose that the time lags longer than
predicted by the AD model are caused by scattering of AD photons by emission-line
clouds that lie $\gtrsim 1$ lt-day from the black hole. In this scenario, supported
by monitoring of emission-line fluxes, the U-band lag is longer because
more distant clouds re-emit the absorbed AD radiation as Balmer continuum photons whose
wavelengths fall within this band.

A qualitative similarity between {\it Swift} observations of 3C~120 and the Seyfert 
galaxies is the smoothness of the UV-optical variations of the latter compared with
the X-ray fluctuations \citep{McHardy2014,Edelson2015,Edelson2017}. Smoother 
UV-optical than X-ray variations in 3C~120 are 
evident when comparing Figures \ref{fig1} and \ref{fig2}. The smoothing time needed to
match the X-ray with the UV time-scales, $\sim 4$ days, is less than the mean X-ray/UVW2 
lag  of 6 days (see \S\S\ref{subsec:cor} and Fig.\ \ref{fig8}). The smoothing time is, 
however, much longer than 
the light-travel time in the UV-optical region in the AD model. We detect no time delay in the radio galaxy between the hard and soft X-ray variations, as seen in NGC4151 
\citep{Edelson2017}. Furthermore, the spectral fits 
to the XRT data of 3C~120 at all epochs are consistent with a single power-law (see
\S\S\ref{subsec:xray} and Fig.\ \ref{fig1}), with no prominent 
soft X-ray excess in 3C~120 in 2016-17. We note that such an excess has been reported 
at earlier epochs; \citet{Lohfink2013} suggested that it could be a synchrotron component 
from the jet. If so, the value of the maximum relativistic electron energy and/or
magnetic field would need to be lower in 2016-17 in order for us not to detect a soft
X-ray excess.

In the cases of NGC5548 and NGC4151, \citet{Gardner2017} and \citet{Edelson2017}
have proposed that the lags correspond to reprocessing of coronal X-ray emission into 
extreme UV (EUV) photons by a hot plasma torus in the innermost portion of the accretion 
flow, plus the reaction time of the AD atmosphere to adjust to changes in the EUV flux. 
The plasma torus emits the soft X-ray excess and mediates the reprocessing of the corona's 
X-rays, with the re-emitted photons heating
the disk at larger radii. The double reprocessing leads to longer time delays between
X-ray and UV-optical variations than predicted by the standard AD model. The lack 
of such a hot torus in 3C~120 in 2016-17, inferred from the absence of a soft X-ray
excess, requires another explanation for the X-ray/UV time lag during dip 2 and the MJD 
57726-57850 flaring episodes, and the smoothness of the UV-optical relative to the
X-ray variations.

A departure of the data from the expectations of the variable AD model for the flares in 
3C~120 proposed in \S\S\ref{subsec:mod} is the X-ray/UVW2 lag of $-10\pm3$ days
during the flaring period (see Fig.\ \ref{fig9}, bottom-left panel). Inspection of the
light curves (Fig.\ \ref{fig2}) reveals that this is mainly due to the variability over 
$\sim 50$ days following flux dip 2. Such long lags could have been caused by a 
disturbance starting in the innermost AD and corona that then propagated outward through 
the AD at a speed much less than $c$,
heating the surface of the AD as it passed through. Such slow (as well as much slower) 
disturbances were inferred to propagate both inward and outward in the study of 
\citet{Chatterjee2009} based on correlations between X-ray and optical variability.
Dip 1 could have been caused by such a disturbance that started in the AD at a radius
somewhat greater than $r_{\rm in}$, 
affecting the UVW2 and UVM2 flux first, then the corona and larger radii in the AD.
However, dip 2 (Fig.\ \ref{fig3}, right) started first with the X-ray decline, 
followed several days later by roughly simultaneous drops, and then recoveries, at all 
UV-optical bands. Instead, we conclude in \S\S\ref{subsec:radio} below that the X-ray and UV-optical flares following dip 2 occurred in the parsec-scale jet.

The shorter-term, relatively low-amplitude X-ray and UV-optical variations seen in
3C~120, with time lags $< 1$ day, can potentially still be
explained by heating of the disk by irradiation from a variable flux of X-rays
generated in the corona \citep{Fabian2015,Fabian2017}. The changes in X-ray flux would
need to heat and cool the disk on time-scales on the order of, or shorter than, the
light-travel time from the corona to the affected area of the disk. 

\begin{figure}[ht!]
\plotone{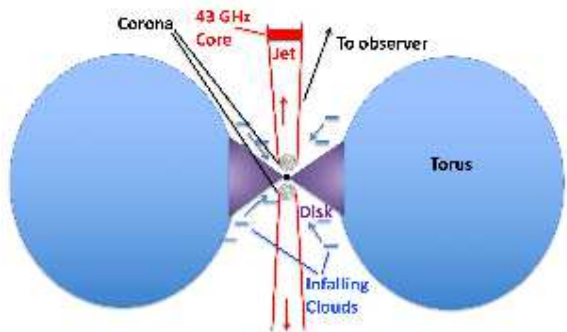}
\vskip 0.5cm \noindent
\caption{Sketch of model (not drawn to scale)
for 3C~120 discussed in the text. The infalling clouds, whose
ionized sides face the AD, scatter radiation from the corona (drawn at the
base of the jet) and disk. The area filling factor of the accretion disk needs to
be $\lesssim 0.5$ for most of the scattered photons to cross the disk plane toward the 
observer without being absorbed. \label{fig12}}
\end{figure}
The complex longer-term, higher-amplitude variations, with time delays that extended up to 
14 days in 2016-17 and temperature variations in excess of that predicted by equation
(1) --- including dip 2 with its long X-ray/UV lag but no significant
UVW2/V lag --- requires one or more additional physical processes that contribute to the 
observed UV-optical flux. Propagating changes in the magnetic field, discussed in
\S\S\ref{subsec:dips}, might be one such mechanism. Another possible effect can be found
within the ``bird's nest'' geometry
of thermal gas in an active nucleus \citep{Mannucci1992,Gaskell2013,Abolmasov2017}.
In this picture, illustrated in Figure \ref{fig12},
the disk flares out into a very broad (polar opening angle
$\lesssim 40^\circ$), clumpy distribution of clouds. The clouds closest to the
symmetry axis (which would presumably be the jet direction in 3C~120) are essentially
falling toward the black hole. If the clouds are optically thick, then they scatter into
our direction light from the AD only on the side facing the disk, hence we observe 
scattered photons mainly from clouds on the far side of the system. This presumes that
the area filling factor of the AD $f \lesssim 0.5$ so that most photons can cross the AD
from the far side to reach the observer. A flare in the corona or inner disk is then both 
observed directly and scattered with a time delay equal to twice the distance from the
inner disk to the scattering cloud. Some of the scattering might be in the Rayleigh
regime, leading to wavelength dependence of the amplitude and time delays of the
flux variations, as in dip 1. Thomson scattering would
produce similar light-curve profiles, as in the UVM2 to V variations of dip 2, which
all decreased by $12\pm1\%$ and varied with little or no cross-frequency lags. In such
a remote scattering geometry, the short-term fluctuations would all be smoothed out,
so that such fluctuations are from emission viewed directly from the AD. Only major,
longer-term variations would be apparent in the scattered light. Clouds located $\sim 3$
lt-days from the inner AD have a free-fall velocity of $\sim 0.05c$ and cross the
region over a time of $\sim 60$ days. This is similar to the time-scales of the flares
seen in the light curves (Fig.\ \ref{fig2}), although we find below that the location of
the flares is in the parsec-scale jet. In order to scatter (with an
albedo $a$) enough radiation to cause a flux increase by a factor $(1+x)$, a disk-shaped
cloud at a radial distance $r$ would need to have a radius
$R \sim (4x/a)^{1/4}(R_{\rm AD}r)^{1/2}$, which results in
$R \sim 7\times10^{14}$ cm = 0.3 lt-days for
$x\sim0.3$, $a\sim0.2$, $R_{\rm AD}\sim2R_{\rm in}\sim2.6\times10^{13}$ cm 
(the radial position in the AD where the photons that are scattered originate), and
$r\sim3$ lt-days. The size requirement is therefore not excessive for such a
scattering event. Furthermore, multiple clouds may be involved at any given time.

\begin{figure}[ht!]
\plotone{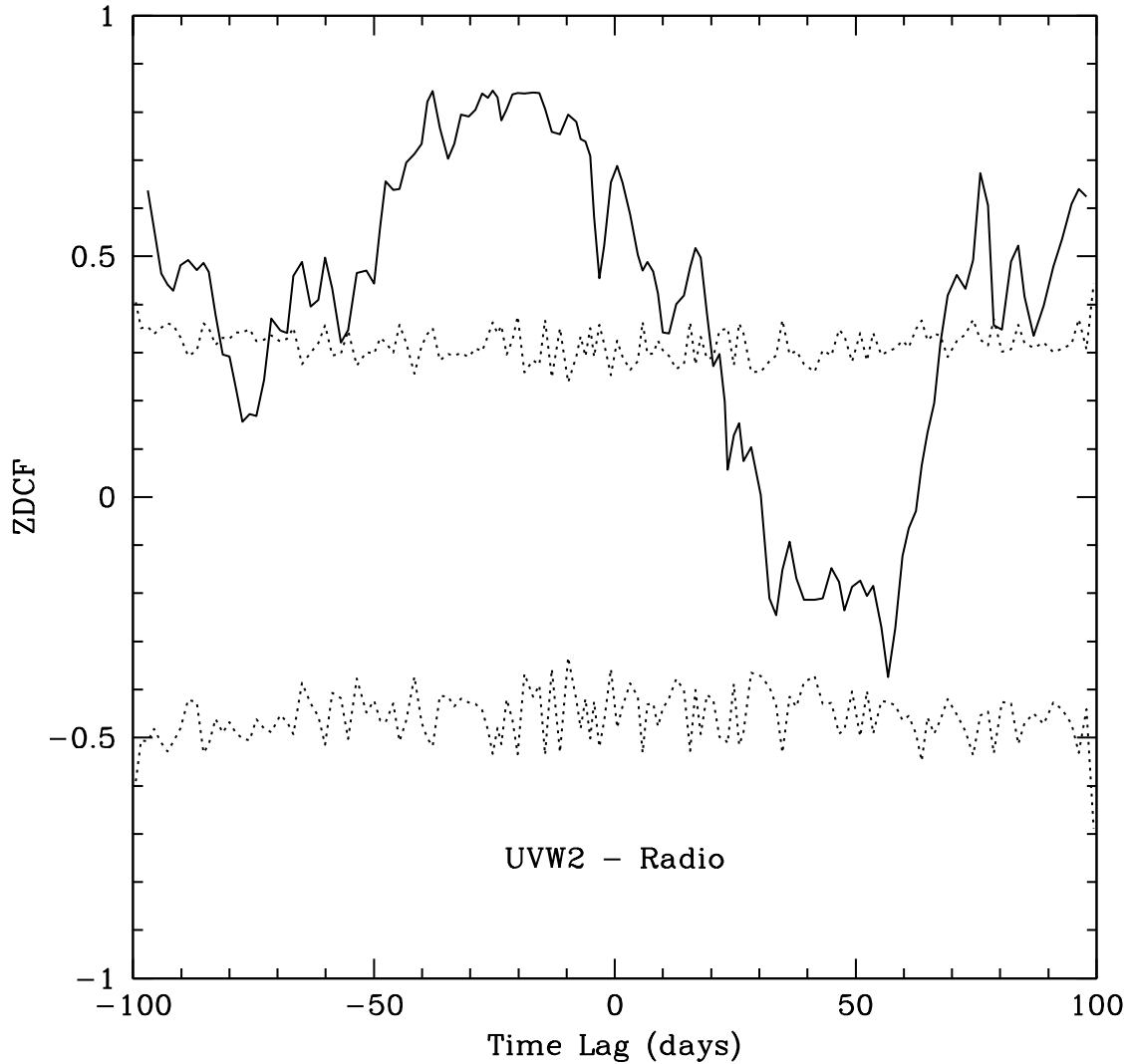}
\vskip 0.5cm \noindent
\caption{Cross-correlation function of UVW2 and 37 GHz variations. Dotted lines
have same meaning as in Figure \ref{fig8}. Negative lags correspond to the first
waveband listed leading the variations.
\label{fig13}}
\end{figure}

\subsection{Correlation of 37 GHz and Optical-UV light Curves: Flares in the Jet} \label{subsec:radio}

A similarity between the 37 GHz and UV-optical light curves, with a radio lag, is apparent 
in Figure \ref{fig4}. The correlation between the UVW2 and 37 GHz variations over
the period of our {\it Swift} monitoring (Fig.\ \ref{fig13}) is strong and highly 
significant statistically, although rather flat at lags from 10 to 40 days, indicating a 
mean delay of 25 days relative to UVW2. (The V/37 GHz correlation is very similar.)
As seen in Figure \ref{fig9}, the mean X-ray/UVW2 
lag during the flaring period was 10 days, for a total X-ray to
37 GHz lag of 35 days. Our sequence of VLBA images demonstrate that the 37 GHz flare 
in 2017 occurred mainly in knots $K3$ and $K4$ (see Fig.\ \ref{fig7}). (Images
available on the Boston U.\ website\footnote{www.bu.edu/blazars/VLBA\_GLAST/3c120.html} 
show that a bright knot ejected earlier was responsible for the radio flare in early 2016, 
which was also preceded by an optical flare; see Fig.\ \ref{fig4}.) Analysis of the 
sequence of VLBA images (Fig.\ \ref{fig6}) indicates that $K4$ was crossing 
quasi-stationary feature $A1$ 0.14 mas downstream of the core during the peak flux 
density of the 37 GHz flare. Given its apparent speed of $1.44\pm0.13$ mas yr$^{-1}$, $K4$
traversed $0.14\pm0.1$ mas over 35 days, and therefore was crossing the core at the
time of the X-ray flare, 0.5-1.3 pc from the black hole (see \S\S\ref{subsec:dips} above).

Since there is no plausible source of thermal radiation at 0.5-1.3 pc with the observed 
inverted spectral slope of $\sim 0.3$ of the flare, we conclude that the UV-optical flare
was synchrotron radiation from a narrow energy distribution of electrons, as discussed
in \S\S\ref{subsec:mod} above. The most likely X-ray emission mechanism is inverse
Compton scattering. The seed photons for the scattering are probably from emission-line 
clouds or hot nuclear dust, since the synchrotron self-Compton process would require the
synchrotron flare to peak prior to the X-ray flare, contrary to the $\sim{-10}$-day
X-ray/UVW2 lag. The decline in available seed photons with distance from the black hole
can explain why the X-ray variations lead those at longer wavelengths.

We note that the optical light curve in \citet{Chatterjee2009} contained three major
flares. Only one of these was temporally associated with an outburst at 37 GHz, 
with a $\sim 1$ month delay relative to the optical event. From this, we conclude that
some, but not all, major optical flares occur in the parsec-scale jet.

Although our optical linear polarization observations (see Fig.\ \ref{fig4}) are sparse, 
some of the measurements were obtained near the peak of the UV-optical flares. The degree of polarization $P$ was between 1.0\% and 2.2\%, with the highest value occurring when 
$\chi$ was nearly parallel to the inner jet direction. This is consistent with turbulent
plasma crossing a standing conical shock, as has been proposed to explain quasi-stationary
features in relativistic jets, such as the core and $A1$ \citep{Marscher2014}.
\citet{Jorstad2007} reported a lower value of
$P\sim0.3\%$ at R-band in 1999-2001, which was consistent with interstellar rather than
intrinsic polarization. The degree of polarization via electron scattering of thermal 
AD emission is expected to be considerably lower than 1\% at optical wavelengths for a 
disk whose pole subtends an angle $\lesssim 20^\circ$ to the line of sight 
\citep{Laor1990,Marin2015}. The electric vector should be transverse to the polar 
direction, and therefore perpendicular to the jet axis. The values of $P$ and $\chi$ that 
we measured during the flare are therefore inconsistent with that expected from the AD.
The low degree of polarization corresponds to a rather disordered magnetic field, as
is commonly found in the 43 GHz cores of compact extragalactic jets \citep{Jorstad2007}. 
This conclusion is supported by the time variability of both the degree and position angle 
of the polarization, which is a sign of turbulence \citep[e.g.,][]{Marscher2014}. The position angles of polarization of the moving knots appear unrelated to those measured in 
R band (cf.\ Figs.\ \ref{fig4} and \ref{fig5}),
hence there is no indication that they produce substantial optical emission when they
are not interacting with either the core or feature $A1$.

Based on 15 GHz VLBA images, \citet{LT2010} found a coincidence between epochs of optical 
flares in 3C~120 and times when a superluminal knot was crossing a stationary emission 
feature $\sim0.7$ mas from the core. Our model fits to the 43 GHz VLBA
data do not include a stationary feature near the same location, although it could be
confused with emission from the superluminal knots, which were ejected more frequently 
than normal in 2016-17. At the peak of the optical flare, there is a local maximum in the 
43 GHz intensity 0.8 mas from the core. However, the feature is very weak ($\sim0.07$ Jy), 
so we consider it unlikely to be the source of the steadier UV-optical synchrotron 
emission inferred in \S\S\ref{subsec:mod} to be present in 3C~120 in 2016-17. 

\section{Summary} \label{sec:summary}

Our monitoring of the radio galaxy 3C~120 at X-ray, UV-optical, and millimeter 
wavelengths over a nine-month time span has revealed that dips in X-ray flux prior to the 
appearance of new superluminal radio knots are accompanied by dips in the UV-optical 
flux as well. Short-term, low-amplitude variations in flux in 3C~120 are well correlated 
across wavelengths, with zero lag to within the uncertainties. An exception is U band,
whose variations lag by $\sim 1$ day, probably from scattering via Balmer continuum
emission in clouds $\sim 1$ lt-day from the black hole. In contrast, longer-term,
more pronounced variations have lags from days to weeks. 

The relative timing of the decline in X-ray and UVW2 flux favors an origin 
in the innermost AD for dip 1 and in the X-ray emitting corona for dip 2. After 
the minimum flux occurs in dip 1, the time for restoration of the flux 
to pre-dip levels is shorter in the UVW1 and UVM2 bands than at U band, contrary to the
\citet{Lohfink2013} scenario in which the inner AD is refilled from larger to smaller
radii. During dip 2, the X-ray flux declines and recovers first, with the UV-optical flux
varying simultaneously within the uncertainties. The time delays of the high-amplitude
variations are generally longer than predicted by a standard AD model. A
scenario that explains such events as the result of re-alignment of the magnetic field to 
expedite flow into the base of the jet, which plays the role of a corona of hot electrons 
\citep{Livio2003,Chatterjee2009}, is more consistent with the observations.
An additional feature, such as the addition of scattering clouds falling toward the black 
hole, is needed to explain the lack of long time delays at longer wavelengths during
dip 2.

A two-component model 
consisting of an inverted-spectrum (IS) source plus a synchrotron source with a spectral 
index of 1.2 can explain the UV-optical spectrum in both quiescent and flaring states. 
The spectral index $\alpha\sim 1/3$ of the IS source is consistent with either thermal 
emission from a standard thin AD or 
synchrotron radiation from a nearly mono-energetic population of electrons of energy 
$\sim10$ GeV. The low, but significant R-band polarization supports the synchrotron model 
and implies that the magnetic field is highly disordered in the emission region.

In contrast to observations of the radio-weak Seyfert galaxy NGC4151 
\citep{Edelson2017}, 3C~120 does not exhibit a soft X-ray excess
whose variations are time-delayed relative to harder X-rays in 2016-17. There is also no 
evidence in the radio galaxy for a hot torus that mediates reprocessing of X-rays from the 
corona to heat more extended regions of the AD \citep{Gardner2017}. Instead, we find that 
the X-ray and UV-optical variations that we have observed in 3C~120 can be attributed to 
physical changes (e.g., magnetic reconfigurations) that propagate at subluminal speeds 
through the inner AD and corona, effects that a varying flux of X-rays from the corona 
have on the inner AD, and scattering of AD emission by hot clouds falling toward the AD at 
a distance of a few lt-days from the black hole. Some longer-term flares, such as those
observed in 2015-17, correspond to nonthermal radiation emitted as a superluminal knot
crosses the 43 GHz core and quasi-stationary feature $A1$ that lies 0.14 mas from the 
core. As the knot crosses the core, a nearly mono-energetic population of electrons
with energies above 10 GeV is added to the power-law distribution at lower energies.
This could result from either magnetic reconnections \citep{Petro2016} or diffusive
shock acceleration in locations where the turbulent magnetic field has an orientation
that is favorable for particle acceleration \citep{Marscher2014}. A similar narrow
energy distribution of electrons at energies $\gtrsim 10$ GeV is required by models
for TeV $\gamma$-ray emission from some BL Lacertae objects \citep[e.g.][]{Balokovic2016}.

Our observations of 3C~120 and those of Seyfert galaxies without strong relativistic
jets reveal AGN to contain extremely complex environments. It appears that nearly all
physical processes that can happen do occur at different times and even different 
locations. The emission that we observe includes photons that arrive to us (and to
scattering regions) both directly and indirectly. Except for the nonthermal flares that 
can erupt on occasion in the parsec-scale jet, there is no obvious major difference in the 
observed X-ray, UV, and optical properties of 3C~120 and Seyfert galaxies as a class. 
Instead, the  physical conditions of the inner AD and its surroundings appear rather
similar. This implies that the difference between AGN with and without powerful jets lies 
in the properties of the black hole and its immediate environs, such as the magnetic flux 
accumulated near the event horizon \citep{TNM2011}.


\acknowledgments

The author thank I.\ McHardy, R. Antonucci, and B. Punsly for informative discussions.
This study was supported in part by NASA through {\it Swift} Guest Investigator grant
NNX16AN69G and {\it Fermi} Guest Investigator grants NNX14AQ58G and 80NSSC17K0649,
and by National Science Foundation grant AST-1615796. This publication makes use of data obtained at the Mets\"ahovi Radio Observatory, operated by Aalto University in Finland.
The VLBA is an instrument of the Long Baseline Observatory. The Long Baseline Observatory 
is a facility of the National Science Foundation operated under cooperative agreement by 
Associated Universities, Inc. This research has made use of the NASA/IPAC Extragalactic Database (NED), which is operated by the Jet Propulsion Laboratory, California Institute of Technology, under contract with the National Aeronautics and Space Administration. 

%

\vspace{5mm}
\facilities{Swift, VLBA}
\software{XSPEC (Arnaud 1996), Difmap (Shepherd 1997), AIPS (van Moorsel, Kemball, \& 
Greisen 1996), HEAsoft (v6.19, 6.21; Arnaud 1996)}




\begin{thebibliography}{}

\bibitem[Abolmasov(2017)]{Abolmasov2017} Abolmasov, P. 2017, \aap, 600, A79
\bibitem[Alexander(1997)]{Alexander1997} Alexander, T.\ 1997, in Astrophysics and Space Science Library 218, Astronomical Time Series, ed.\ D.\ Maoz, A.\ Sternberg, \& E.\ M.\ 
Leibowitz,  163
\bibitem[Alexander(2013)]{Alexander2013} Alexander, T.\ 2013, arXiv:1302.1508
\bibitem[Arnaud(1996)]{Arnaud1996} Arnaud, K.\ A.\ 1996, in ASP Conf. Ser. 101, 
Astronomical Data Analysis Software and Systems V, ed.\ G.\ Jacoby \& J.\ Barnes, 17
\bibitem[Balokovic et al.(2016)]{Balokovic2016} Balokovic, M., Paneque, D., Madejski,
G., et al. 2016, \apj, 819, 156
\bibitem[Barvainis(1993)]{Barvainis1990} Barvainis, R.\ 1993, \apj, 412, 513
\bibitem[Bentz \& Katz(2015)]{Bentz2015} Bentz, M.\ C., \& Katz, S.\ 2015, \pasp, 127, 67
\bibitem[Blandford \& Payne(1977)]{BP1982} Blandford, R.\ D., \& Payne, D.\ G.\ 1982, \mnras, 199, 883
\bibitem[Blandford \& Znajek(1977)]{BZ1977} Blandford, R.\ D., \& Znajek, R.\ 1977, \mnras, 179, 433
\bibitem[Buisson et al.(2017)]{Buisson2017} Buisson, D.\ J.\ K., Lohfink, A.\ M., Alston,
W.\ N., \& Fabian, A.\ C. 2017, \mnras, 464, 3194
\bibitem[Cackett et al.(2018)]{Cackett2018} Cackett, E.\ M., Chiang, C.-Y., McHardy,
I.\ M., et al. 2018, \apj, 857, 53
\bibitem[Cackett, Horne, \& Winkler(2007)]{Cackett2007} Cackett, E.\ M., Horne, K., 
\& Winkler, H. 2007, \mnras, 380, 669
\bibitem[Casadio et al.(2015)]{Casadio2015} Casadio, C., G\'omez, J.\ L., Grandi, P., et al.\ 2015, \apj, 808, 162
\bibitem[Cawthorne(2006)]{Cawthorne2006} Cawthorne, T.\ V.\ 2006, \mnras, 367, 851
\bibitem[Cawthorne, Jorstad, \& Marscher(2013)]{Cawthorne2013} Cawthorne, T.\ V.\ 2013, \apj, 772, 14
\bibitem[Chatterjee et al.(2009)]{Chatterjee2009} Chatterjee, R., Marscher, A.\ P., Jorstad, S.\ G., et al.\ 2009, \apj, 704, 1689 
\bibitem[Chatterjee et al.(2011)]{Chatterjee2011} Chatterjee, R., Marscher, A.\ P., Jorstad, S.\ G., et al.\ 2011, \apj, 734, 43
\bibitem[Denn, Mutel, \& Marscher(2000)]{Denn2000} Denn, G.\ R., Mutel, R.\ L., \&
Marscher, A.\ P. 2000, \apjs, 129, 61
\bibitem[Dickey \& Lockman(1990)]{Dickey1990} Dickey, J.\ M., \& Lockman, F.\ J.\  
1990, \araa, 28, 215
\bibitem[Edelson et al.(2015)]{Edelson2015} Edelson, R., Gelbord, J.\ M., Horne, K., et al.\ 2015, \apj, 806, 129
\bibitem[Edelson et al.(2017)]{Edelson2017} Edelson, R., Gelbord, J.\ M., Cackett, E., et al.\ 2017, \apj, 840, 41
\bibitem[Eracleous, Sambruna, \& Mushotzky(2000)]{Eracleous2000} Eracleous, M., Sambruna, R., \& Mushotzky, R.\ F.\ 2000, \apj, 537, 654
\bibitem[Fabian et al.(2015)]{Fabian2015} Fabian, A.\ C., Lohfink, A., Kara, E., et al.\ 2015, \mnras, 451, 4375
\bibitem[Fabian et al.(2017)]{Fabian2017} Fabian, A.\ C., Lohfink, A., Belmont, R.,
Malzac, J., \& Coppi, P.\ 2017, \mnras, 467, 2566
\bibitem[Fanaroff \& Riley (1974)]{FR1974} Fanaroff, B.\ L., \& Riley, J.\ M.\ 1974, \mnras, 167, 31P
\bibitem[Fender, Belloni, \& Gallo(2004)]{Fender2004} Fender, R.\ P., Belloni, T.\ M., \& Gallo, E.\ 2004, \mnras, 355, 1105
\bibitem[Fitzpatrick(1999)]{Fitzpatrick1999} Fitzpatrick, E.\ L. 1999, PASP, 111, 63
\bibitem[Gardner \& Done(2017)]{Gardner2017} Gardner, E., \& Done, C.\ 2017, \mnras, 470, 3591
\bibitem[Gaskell \& Goosmann(2013)]{Gaskell2013} Gaskell, C.\ M., \& Goosmann, R.\ W.\ 
2013, \apj, 769, 30
\bibitem[Gelbord et al.(2015)]{Gelbord2015} Gelbord, J., Gronwall, C., Grupe, D.,
Vanden Berk, D., \& Wu, J.\ 2015, in Swift: 10 Years of Discovery, ed. P.\ Caraveo, P.\  
D'Avanzo, N.\ Gehrels, \& G.\ Tagliaferri, PoS Swift, 10, 137
\bibitem[G\'omez et al.(2001)]{Gomez2001} G\'omez, J.\ L., Marscher, A.\ P., Alberdi, A., Jorstad, S.\ G., \& Agudo, I.\ 2001, \apjl, 561, L161
\bibitem[G\'omez et al.(2011)]{Gomez2011} G\'omez, J.\ L., Roca-Sogorb, M., Agudo, I., Marscher, A.\ P., \& Jorstad, S.\ G.\ 2001, \apj, 733, 11
\bibitem[Jorstad \& Marscher(2016)]{Jorstad2016} Jorstad, S.\ G., \& Marscher, A.\ P.\  2016, Galaxies, 4, 47
\bibitem[Jorstad et al.(2005)]{Jorstad2005} Jorstad, S.\ G., Marscher, A.\ P., Lister, M.\ L., et al.\ 2005, \aj, 130, 1418
\bibitem[Jorstad et al.(2007)]{Jorstad2007} Jorstad, S.\ G., Marscher, A.\ P., Stevens, J.\ A., et al.\ 2007, \aj, 134, 799
\bibitem[Jorstad et al.(2010)]{Jorstad2010} Jorstad, S.\ G., Marscher, A.\ P., Larionov, 
V.\ M., et al. 2010, \apj, 715, 362
\bibitem[Jorstad et al.(2017)]{Jorstad2017} Jorstad, S.\ G., Marscher, A.\ P., Morozova, 
D.\ A., et al. 2017, \apj, 846, 98
\bibitem[King, Lohfink, \& Kara(2017)]{King2017} King, A.\ L., Lohfink, A., \& Kara, E.
2017, \apj, 835, 226
\bibitem[Kishimoto et al.(2008)]{Kishimoto2008} Kishimoto, M., Antonucci, R., Blaes, 
O., et al.\  2008, \nat, 454, 492
\bibitem[Lampton, Margon, \& Bowyer (1976)]{Lampton1976} Lampton, M., Margon, B., \&
Bowyer, S.\  1976, \apj, 208, 177
\bibitem[Laor, Netzer, \& Piran(1990)]{Laor1990} Laor, A., Netzer, H., \& Piran, T.\ 1990,
\mnras, 242,560
\bibitem[Le\'on-Tavares et al.(2010)]{LT2010} Leon-Tavares, J., Lobanov, A.\ P.,
Chavushyan V.\ H., et al.\ 2010, \apj, 715, 355
\bibitem[Lira et al.(2011)]{Lira2011} Lira, P., Ar\'evalo, P., Uttley, P., McHardy, I.,
\& Breedt, E. 2011, \mnras, 415, 1290
\bibitem[Livio, Pringle, \& King(2003)]{Livio2003} Livio, M., Pringle, J.\ E., \& King, A.\ R.\ 2003, \apj, 593, 184
\bibitem[Lohfink et al.(2013)]{Lohfink2013} Lohfink, A.\ M., Reynolds, C.\ S., Jorstad, S.\ G., et al.\ 2013, \apj, 772, 83
\bibitem[Lyutikov, Pariev, \& Gabuzda(2005)]{Lyut05} Lyutikov, M., Pariev, V. I., \& Gabuzda, D. C. 2005, \mnras, 360, 869
\bibitem[Mannucci, Salvati, \& Stanga(1992)]{Mannucci1992} Mannucci, F.,
Salvati, M., \& Stanga, R.\ M.\ 1992, \apj, 394, 98
\bibitem[Marin, Goosmann, \& Gaskell(2015)]{Marin2015} Marin, F., Goosmann, R.\ W., \& Gaskell, C.\ M.\ 2015, \aap, 577, A66
\bibitem[Markoff, Nowak, \& Wilms(2005)]{Markoff2005} Markoff, S., Nowak, M.\ A., \& Wilms, J.\ 2005, \apj, 635, 1203
\bibitem[Marscher (2014)]{Marscher2014} Marscher, A.\ P.\  2014, \apj, 780, 87
\bibitem[Marscher et al.(2002)]{Marscher2002} Marscher, A.\ P., Jorstad, S.\ G., G\'omez, 
J.\ L., et al.\  2002, \nat, 417, 625
\bibitem[Marscher, Jorstad, \& Williamson(2017)]{Marscher2017} Marscher, A.\ P.,
Jorstad, S.\ G., \& Williamson, K.\ E.\ 2017, Galaxies, 5, 63
\bibitem[Marshall et al.(2009)]{Marshall2009} Marshall, K., Ryle, W.\ T., Miller, H.\ R., et al.\ 2009, \apj, 696, 601
\bibitem[McHardy et al.(2014)]{McHardy2014} McHardy, I.\ M., Cameron, D.\ T., Dwelly, T.,
et al.\ 2014, \mnras, 444, 1469
\bibitem[McHardy et al.(2018)]{McHardy2018} McHardy, I.\ M., Connolly, S.\ D., Horne, K.,
et al.\ 2018, \mnras, 480, 2881
\bibitem[McKinney \& Narayan(2007)]{MN2007} McKinney, J.\ C., \& Narayan, R.\  2007, \mnras, 375, 513
\bibitem[Ogle et al.(2005)]{Ogle2005} Ogle, P.\ M., Davis, S.\ W., Antonucci, R.\ R.\ J., et al.\ 2005, \apj, 618, 139
\bibitem[Pacholczyk(1970)]{Pach1970} Pacholczyk, A.\ G.\ 1970, Radio Astrophysics
(San Francisco: Freeman), p.\ 90
\bibitem[Peterson et al.(1998)]{Peterson1998} Peterson B.\ M., Wanders I., Horne K., et al. 1998, \pasp, 110, 660
\bibitem[Petropoulou, Giannios, \& Sironi(2016)]{Petro2016} Petropoulou, M., Giannios, D., \& Sironi, L. 2016, \mnras, 462, 3325
\bibitem[Punsly et al.(2015)]{Punsly2015} Punsly, B., Marziani, P., Kharb, P., O'Dea, C.\ P., \& Vestergaard, M.\ 2015, \apj, 812, 79
\bibitem[Rani \& Stalin(2018)]{Rani2018} Rani, P., \& Stalin, C.\ S.\  2018, \apj, 856, 120
\bibitem[Schlafly \& Finkbeiner(2011)]{SF11} Schlafly, E.\ F., \& Finkbeiner, D.\ P.
2011, ApJ, 737, 103
\bibitem[Shakura \& Sunyaev(1973)]{Shakura1973} Shakura, N., \& Sunyaev, R.\ 1973, \aap, 24, 337
\bibitem[Shepherd(1997)]{Shepherd1997} Shepherd, M.\ C.\ 1997, in ASP Conf. Proc. 125, 
Astronomical Data Analysis Software and Systems VI, ed.\ G.\ Hunt \& H.\ E.\ Payne
(San Francisco, CA: ASP), 77
\bibitem[Tchekhovskoy, Narayan, \& McKinney(2011)]{TNM2011} Tchekhovskoy, A., Narayan, R., \& McKinney, J.\ C.\ 2011, \mnras, 418, L79
\bibitem[Ter\"asranta et al.(1998)]{Terasranta1998} Ter\"asranta, H., Tornikoski, M.,
Mujunen, A., et al. 1998, \aaps, 132, 305
\bibitem[van Moorsel, Kemball, \& Greisen(1996)]{vanMoorsel1996} van Moorsel, G., Kemball, 
A., \& Greisen, E.\ 1996, in Astronomical Data Analysis Software and Systems V, ASP Conf. Ser., 101, ed.\ G.\ H.\ Jacoby \& J.\ Barnes, 37
\bibitem[Vlahakis \& K\"onigl(2004)]{VK2004} Vlahakis, N., \& K\"onigl, A.\  2004, \apj, 605, 656
\bibitem[Williamson et al.(2014)]{Williamson2014} Williamson, K.\ E., Jorstad, S.\ G.,
Marscher, A.\ P., et al. 2014, \apj, 789, 135
\bibitem[Williamson et al.(2016)]{Williamson2016} Williamson, K.\ E., Jorstad, S.\ G.,
Marscher, A.\ P., et al. 2016, Galaxies, 4, 64

\end{thebibliography}
\end{document}